\documentclass{elsart}

\usepackage{amssymb,epsfig}
\newcommand{\beq}{\begin{equation}}
\newcommand{\eneq}{\end{equation}}
\newcommand{\vpl}{v_{\rm pl}}
\begin{document}

\begin{frontmatter}



\title{Boundary Field Theory Approach to the Renormalization of SQUID Devices}


\author{Domenico Giuliano$^1$  and Pasquale Sodano$^2$}

\address{$^1$ Dipartimento di Fisica, Universit\`{a} della Calabria and
I.N.F.N., Gruppo collegato di Cosenza \\ Arcavacata di Rende
I-87036, Cosenza, Italy\\
$^2$ Progetto Lagrange- Fondazione C.R.T. and Fondazione I.S.I.,\\
Dipartimento di Fisica, Politecnico di Torino,\\ Corso Duca degli
Abruzzi 24, Torino, Italy \thanksref{Perugia}}

\thanks[Perugia]{Permanent address: Dipartimento di Fisica, Universit\`{a} di
Perugia, and I.N.F.N., Sezione di Perugia, Via A. Pascoli, 06123,
Perugia, Italy}

\begin{abstract}

We show that the quantum properties of some Josephson SQUID devices are
described by a boundary sine Gordon model.  Our approach naturally
describes multi-junction SQUID devices and, when applied to a single
junction SQUID (the rf-SQUID), it reproduces the known results of
Glazman and Hekking. We provide a detailed analysis of the regimes
accessible to an rf-SQUID and to a two-Josephson junction SQUID
device (the dc-SQUID). We then compute the normal component of the
current-response of a 
SQUID device to an externally applied voltage and show that the 
equation describing the current-voltage characteristic function
reduces to well-known results when the infrared cutoff is suitably chosen.
Our approach helps in establishing new and interesting
connections between superconducting devices, quantum brownian
motion, fermionic quantum wires and, more generally, quantum
impurity problems.

\end{abstract}

\end{frontmatter}


\section{Introduction}

Quantum effects in Josephson junctions have been by now investigated
for quite a long time, both experimentally \cite{devoret} and
theoretically \cite{schon}: the most important manifestation of
quantum fluctuations in these systems being the macroscopic quantum
tunneling of the phase across a current biased junction and the
consequent observation of quantum phase slips events at a bias
rather close to the Josephson critical current \cite{devoret}.

Early studies of quantum phase fluctuations in Josephson systems
have been carried using either the Caldeira- Leggett \cite{caldeira}
or the electromagnetic environment \cite{em} models: within this
framework, it has been shown that macroscopic quantum tunneling
causes a finite voltage to appear for any finite current, leading to
a non-linear I-V characteristic which, in the limit of zero
temperature, reads as $V=AI^{\gamma}$ with ${\gamma}$ depending on
the impedance of the superconducting leads and $A$ being
proportional to the bare tunnel matrix element between two adjacent
minima of the potential \cite{GraFisher}. The dual result for a
voltage biased junction shows that the dc-current in such a junction
is proportional to the square of the unrenormalized Josephson energy
\cite{AveFaIn}.

As already   pointed out in Ref.\cite{glazhek}, in the above
mentioned approaches the effective boundary conditions for the
quantum fluctuations of the "environment modes" do not depend on the
Josephson energy of the junction. While this assumption is perfectly
legitimate for weak fluctuations of the phase of  the order
parameter across the junction, better care should be used if these
fluctuations are strong as it may well happen for one-dimensional
superconductors, where phase fluctuations diverge logarithmically
with the length of the system, opening the undesirable option that
the Josephson energy could average to zero for diverging random
fluctuations of the phase.

In their seminal paper Glazman and Hekking \cite{glazhek} showed
that a finite renormalized Josephson energy may arise since the
junction itself affects the fluctuations of the environment. They
considered an rf-SQUID, i.e. a thin superconducting loop containing
a Josephson junction, phase-biased by threading a flux $\Phi_{\rm ext}$
through the loop. The environment modes of the loop consist of
plasmon modes propagating through a superconducting wire with  a
soundlike dispersion law \cite{schon2}. Their analysis was based on
a perturbative renormalization group approach exploiting the
similarity of the effective action, emerging from integrating out
the fluctuations "away" from the junction, with the one used to
analyze the quantum brownian motion in a periodic potential
\cite{zwergerkane}. They found how the Josephson coupling $E_J$ is
renormalized when high energy degrees of freedom are integrated out
and showed that the renormalization leads to values of the Josephson
critical current which are smaller than one would expect from the
mean field theory result $J^0_c={\pi}{\Delta}G/2e$, where $G$ is the
conductance of the junction and ${\Delta}$ is the superconducting
gap in the loop. The RG approach is supplemented by an instanton
analysis \cite{phaseslips} needed to compute the effect of the
macroscopic quantum tunneling on the phase dependence of the
Josephson current for relatively large loops.

Generalized SQUID devices are much 
studied today mainly because of
their relevance for the implementation of flux qu-bits
\cite{makhlin}: it is by now widely accepted that multi-junction
SQUID devices \cite{QC1} are the most promising realization of a
flux qu-bit due to their reduced size, which renders them less
sensitive to decoherence effects \cite{deco} induced by the
interaction with the environment. As it is well known,
an rf-SQUIDs \cite{QC2} may be represented as a two level system only
when the superconducting loop is rather large and, thus,
unfortunately more sensitive to interactions with the environment.

In this paper we evidence that all the non-perturbative  renormalization
effects induced by quantum fluctuations of the phase of the
superconducting order parameter  \cite{glazhek} are described by
 the two-boundary sine Gordon model \cite{twoboundary}, which
becomes then the pertinent description of the dynamics of the fluctuation
field and of a Josephson junction with arbitrary nominal strength $E_J$.
As we shall see, the quantum behaviors of  rf-SQUIDs and of multi-junction 
SQUID devices, may be analyzed within the framework of boundary field theories
\cite{cardy}.  
 
A two boundary sine Gordon model description of SQUID devices is
possible since the dynamics of the fluctuations of the phase of the
superconducting wire may be described as a spinless Tomonaga-Luttinger liquid
(TLL) \cite{TLL}. In fact, for a superconducting wire of length $ 2 L$ and
cross-sectional area $S$, the Lagrangian describing the phase
fluctuations of the order parameter may be written as
\cite{glazhek}:

\beq L = \int_{-L}^L \: d x \: \left\{ \frac{\hbar^2}{2 e_c} \left(
\frac{\partial \Phi }{\partial t } \right)^2 - \frac{\hbar^2 n_s S}{
4 m } \left( \frac{\partial \Phi}{\partial x} \right)^2 \right\}
\:\:\:\: , \label{eq1.1} \eneq \noindent

where $n_s$ is the superfluid density, $m$ is the electron mass, $1/
e_c$ is the characteristic inverse charging energy per unit length
of the loop:

\beq \frac{1}{e_c} = \frac{\epsilon}{ 8 e^2 \ln ( R / a ) } \;\;\;\:
, \label{eq1.2} \eneq \noindent where $\epsilon$ is the dielectric
constant of the medium surrounding the loop, while $R$ is the
distance from a metallic screen, which, together with $a$, fixes the
lead's capacitance.

Upon setting $\hbar = 1$, if one defines the "Luttinger" parameter
$g = \frac{\pi}{2} \sqrt{n_s S / ( m e_c ) }$ and the plasmon velocity $\vpl =
\sqrt{ n_s S e_c / (4 m) }$, one immediately sees that the
Lagrangian (\ref{eq1.1}) may be written as the one describing a
spinless Luttinger liquid \cite{schulz}:

\beq L = \frac{g }{2 \pi} \int_{-L}^L \: d x \: \left\{
\frac{1}{\vpl} \left( \frac{\partial \Phi}{\partial t} \right)^2 -
\vpl \left( \frac{\partial \Phi}{\partial x} \right)^2 \right\}
\:\:\:\: . \label{eq1.3} \eneq \noindent

\footnote{Our normalization of $g$ renders the boundary interaction
marginal when $g=1$. It differs by the one used in Ref.\cite{wires}
by a factor of 1/2.}

Physically, $1/g$ is the dimensionless zero-frequency impedance of
the superconducting wire \cite{glazhek,matglaz}. Due to its
dependence on $S$, $g$ may take values in a rather wide interval of
real numbers: for $g > 1$, Eq.(\ref{eq1.3}) describes a TLL with an
attractive interaction, while, for $g < 1$ the interaction is
repulsive. The value $g=1$ describes the very special case of an
essentially free theory. Different superconducting wires may realize
different values of $g$: for instance, for rather clean aluminum
wires, one can easily attain values of $g$ such that $g>1$ and,
thus, realize a wire for which the dynamics of the phase
fluctuations is described by an attractive TLL.\cite{glazhek}.

For $g>1$, the two boundary sine Gordon model description of a SQUID
device allows to make very general statements regarding the regimes
accessible to the system for a finite loop-size $L$. Namely, there
will be a perturbative (in $E_J$) weak coupling regime accessible
when the relevant Josephson couplings are small and a
non-perturbative strong coupling regime accessible when $E_J$
becomes large; most importantly, there will be a renormalization
group invariant length scale $L_*$ such that for $L < L_*$ the SQUID
device is in the perturbative weak coupling regime, while it is in
the non-perturbative strong coupling regime for $L > L_*$.
Intuitively speaking, $L_*$ is ($ 2 \pi$ times) the radius for which
the ratio between the Josephson energy and the magnetic energy due
to the loop self-inductance equals one; the existence of an "healing
length" $L_*$ is generic to a boundary sine Gordon model describing
a TLL \cite{GiuSo}.  At variance, for $g<1$, the theory lies within
the perturbative regime at any $L$. Therefore, there is no healing
length in this case and no universal behavior is then attainable. In
the following, we shall show that  these properties are shared by a variety of
SQUID devices.

We investigate also the stability of the two (weakly and strongly
coupled) regimes by looking at the relevance/irrelevance of the
operators describing the boundary interactions. It turns out that
these operators are either vertex operators or instantons (phase
slips) of the plasmon field $\Phi$, depending on if the system is
within the weakly, or the strongly coupled regime. As evidenced in
the following, vertex (instantons) are a relevant (irrelevant)
perturbation of the weakly (strongly) coupled regime for $g>1$,
while they are irrelevant (relevant) for $g<1$.

It emerges the picture that, for $g>1$, it always exists an healing
length $L_*$ such that, for $L > L_*$, the Josephson coupling
affects the dynamics of the plasmon field as a boundary condition:
in this regime it is impossible to disentagle the Josephson junction
from the environmental modes. For $g<1$, instead, this situation is
never attainable and the Josephson coupling has to be always
regarded as a quantum impurity embedded in a superconducting wire.

The strongly and weakly coupled regimes may be probed by looking either at 
the dc-conductance, or at the behavior of the Josephson current as a
function of the magnetic flux treading the superconducting loop.
The two-boundary  sine Gordon approach provides a systematic and
powerful method to evaluate the relevant currents. In fact, the correlators 
of the plasmon fluctuation  $\Phi$ field(s)  may be exactly computed and 
this allows  for a systematic derivation
of the leading corrections to the Josephson current for all values of
the applied flux $\varphi = \Phi_{\rm ext} / \Phi_0^*$ ( $\Phi_0^* = h /
( 2 e )$), which, for $\varphi = \pi$, yields  the results of
the instanton analysis carried in Ref.\cite{glazhek}; in addition, we shall
show that it also allows for a systematic derivation  of the response of a 
SQUID device to an applied external voltage. Since applying a finite voltage
$V$ introduces a new energy scale $e^* V$, with $e^* = 2 e$,
the dc-current  response of the system strongly depends on the ratio between 
the new energy scale, and the intrinsic scale provided by the level spacing 
$s = 2 \pi \vpl / L$ of the plasmon field(s) $\Phi$.
Our computation of the dc-current clearly shows that, when $e^* V > s$, 
one gets the well known Kane-Fisher formula \cite{zwergerkane}, 
while, for $s> e^* V$, one gets the results obtained by Glazman and Hekking  
\cite{glazhek}; in addition, when $e^* V > s$, the dc-current dominates
over the Josephson current.

Boundary field theories have by now become relevant in several
different contexts. In condensed matter theory , they are mostly
generalizations of quantum impurity models, which may be described
by using the TLL paradigm \cite{TLL};
for instance, boundary interactions appear in the analysis of the
Kondo problem \cite{Kondo}, in the study of a one-dimensional
conductor in presence of an impurity \cite{odcimp}, in the
derivation of tunneling between edge states of a Hall bar
\cite{hall}, in the study of quantum brownian motion in a periodic
potential \cite{brownian} and in the analysis of the phases
accessible to networks of quantum wires \cite{wires} and Josephson
junctions \cite{jj},\cite{GiuSo}. The TLL paradigm evidences that
many interactions are simply diagonalizable in the basis of
pertinent collective bosonic modes, and that non-diagonalizable
interactions correspond to exactly solvable Hamiltonians, such as
sine Gordon models \cite{sgmodels}. Recently, boundary field
theories have been investigated in the context of string theories.
For instance, in studying tachyon instabilities \cite{tachyon}, one
is faced with the fact that the space of interacting string theory
\cite{sen} is mapped onto the space of boundary perturbations of
conformal field theories \cite{leclair}, and that the
renormalization group flow determined by boundary perturbations may
be identified with tachyon condensation \cite{narain}. Affleck and
Ludwig \cite{Afflud} showed that the boundary entropy $g$ is
decreasing along the renormalization group trajectories, triggered
by the boundary interactions. It is remarkable that this entropy has
been proposed as a measure of entanglement for boundary states
\cite{Afflud}.

The paper is organized as follows:

Section 2 shows that quantum behaviors of an rf-SQUID are described 
by a two-boundary sine Gordon model.  There, we shall analyze the weak
and the strong coupling regime accessible to
the device for a finite size of the superconducting loop. We show that, 
for $g>1$, there is a renormalization group invariant scale -
the "healing length" $L^*$- \cite{glazhek}  marking 
the crossover between the weakly coupled and the strongly coupled regime.

We first analyze the weakly coupled fixed point
corresponding to mixed boundary conditions (i.e. Dirichlet at the
inner boundary and Neumann at the outer boundary) and we determine
the scale dimension of the pertinent boundary operators and the
scaling equations for the  running coupling constants using
renormalization group (RG) methods; this allows for a perturbative
computation of the beta-function for all the couplings. Computation
of the canonical ground state energy shows that- at the weakly
coupled fixed point- the Josephson current has a sine dependence on
the flux treading the loop. Then, we analyze the strong coupling
fixed point, corresponding to Dirichlet boundary conditions at the
outer and inner boundaries of the rf-SQUID. Evaluation of the
canonical ground state energy shows that- at the strongly coupled
Dirichlet fixed point- the Josephson current exhibits  a saw-tooth
dependence on the external flux.

Due to the zero modes, appearing in the strong coupling regime as a
consequence of the Dirichlet boundary conditions, the ground state is 
doubly degenerate when $\varphi  =\pi+2\pi k$: when these levels are
confined away from the rest of the spectrum, the rf-SQUID realizes a
qu-bit (See, for instance, Ref.\cite{makhlin}). 
Our analysis points out that this realization is possible
only for $g>1$ and for sizes of the superconducting loop bigger than
the healing length, and, thus, for a rather big self-inductance of
the device. Finally, we analyze the effects of quantum phase slips
on the ground state degeneracy of the rf-SQUID at the strong
coupling fixed point: an explicit instanton computation shows that,
for $g>1$, the degeneracy is lifted by an amount proportional to the
instanton fugacity, and that the instantons 
smoothen the edges of the saw-tooth phase-current relation.

In section 3  we analyze a SQUID-device involving only two
Josephson junctions, of arbitrary strength $E_{J,1}$, $E_{J,2}$ (the dc-SQUID).
 Under the assumption that the two plasmon
fields living in the two branches of the superconducting loop (see
fig.1) are described by TLLs corresponding to the same $g$, we 
show that also the dc-SQUID is described by a two-fields generalization 
of the two boundary sine Gordon model. 
We show that also here, for $g>1$, there is an unstable weak coupling
fixed point and a stable one at strong coupling and that the
crossover from one to the other is characterized by an healing
length $L^*$, whose value depends, of course,  only on the smallest of the two
Josephson couplings. When $g<1$, the only stable fixed point is at
weak coupling.

In section 4 we analyze the normal component of the 
current-response of the rf- and dc-SQUIDs to 
an applied external voltage. There, we evidence that a  computation of
the time-independent 
current using the $\Phi$-field(s) correlators derived from the two-boundary
sine Gordon model yields the well-known Kane-Fisher formula when
the infrared cutoff  $e^* V > s$
 \cite{zwergerkane}, as well as
the Glazman-Hekking result \cite{glazhek},  when the infrared cutoff
if $e^* V < s $. It is remarkable that, as a result of the
boundary field theory approach to superconducting SQUID devices developed
in this paper, one gets a current-voltage equation 
valid for any choice of the infrared cutoff. 
Furthermore, for a dc-SQUID, the boundary field theory approach evidences  
the possibility
of new remarkable interference effects (even in the absence of an external 
applied magnetic flux (!)) between the currents flowing across the two 
junctions.

Section 5 is devoted to our conclusions and final remarks.

In order to be pedagogical and self-contained, a few pertinent
mathematical details are summarized in the appendices.

\section{The rf-SQUID}

From Eq.(\ref{eq1.3}) one may easily obtain the Hamiltonian $H_{rf}$
for an rf-SQUID (i.e. for a superconducting wire cut by a Josephson
junction with nominal Josephson coupling $E_J$ and pierced by a
magnetic flux $\varphi$). By neglecting the charging energy of the junction
\cite{glazhek}, $H_{rf}$ is given by:

\[ H_{rf} \equiv H_{0} + H_B = \frac{g}{2 \pi} \int_{-L}^L \: d x \: \left[
\frac{1}{\vpl} \left( \frac{\partial \Phi ( x  ) }{ \partial t}
\right)^2 + \vpl \left( \frac{ \partial \Phi ( x  ) }{ \partial x} -
\frac{ \varphi}{2 L} \right)^2 \right]
\]

\beq
 - E_J \cos [ \Phi ( L ) -
\Phi ( - L ) ] \:\:\:\: , \label{for1} \eneq
where $H_0$ and $H_B$ describe the bulk and boundary Hamiltonian, respectively.

The flux $\varphi$ may be accounted for by a redefinition of the
field $\Phi$ as  $\Phi \to \Phi - \varphi \frac{x}{2L}$. 
By introducing the symmetric and the antisymmetric
phase fields $\Phi_e ( x  ) = \frac{1}{ \sqrt{2}} [ \Phi ( x  ) +
\Phi (  - x ) ] $ and $\Phi_o ( x  ) = \frac{1}{\sqrt{2}} [ \Phi ( x
) - \Phi ( - x  ) ]$, with $0 \leq x \leq L$, one gets

\beq
H_{rf} = \sum_{ j = e , o } \frac{g}{ 2 \pi} \int_0^L \: d x \:
\left[ \frac{1}{\vpl} \left(
\frac{\partial \Phi_j ( x  ) }{ \partial t} \right)^2 +
\vpl \left( \frac{ \partial \Phi_j ( x  ) }{ \partial x} \right)^2 \right]
- E_J \cos [ \sqrt{2} \Phi_o ( L ) + \varphi ]
\:\:\:\: .
\label{for3}
\eneq
The field $\Phi_e (x  )$ fully decouples from the interaction term in
Eq.(\ref{for3}). The Hamiltonian for the field $\Phi_o ( x  )$
(which will be referred to as $\Phi ( x  )$, from now on), on the
other hand, is the Hamiltonian for a two-boundary sine Gordon model
\cite{twoboundary}, with Dirichlet boundary condition at the inner
boundary, i.e., $\Phi ( 0  , t ) = 0$, and dynamical boundary
conditions at the outer boundary. The latter are  obtained by
requiring that the energy functional is conserved, yielding \cite{GiuSo}

\beq
 \frac{g \vpl}{ \pi} \frac{ \partial \Phi ( L , t ) }{ \partial x} +
\sqrt{2}   E_J \sin [  \sqrt{2} \Phi ( L , t )  + \varphi ]  = 0
\:\:\:\: .
\label{eq1.6}
\eneq
\noindent

For $E_J \to 0$, one has Neumann boundary conditions at the inner
boundary, i.e., $  \frac{ \partial \Phi ( L , t ) }{ \partial x} =
0$, while, for $E_J \to \infty$, one gets Dirichlet boundary
conditions, $ \sqrt{2} \Phi ( L , t ) = - \varphi + 2 \pi k$, $k \in
Z$. As it will be clearer in the following, the effective
two-boundary sine Gordon model provides the relevant
renormalizations of an rf-SQUID: namely, it gives  the renormalized
value ($\bar{E}_J$) of the Josephson energy and yields the correct
functional dependence of the Josephson current, as  a function of
the flux $\varphi$.

As we shall see in the following sections, as $L \to \infty$, there are two 
relevant fixed points, namely, the
weakly coupled Neumann fixed point ($E_J = 0$), and the strongly
coupled Dirichlet fixed point ($E_J = \infty$). Our analysis will evidence
the existence of  an healing length $L_* = g \pi \vpl / E_J$, 
providing the size of a loop for which the ratio between the
Josephson energy and the magnetic energy due to the loop
self-inductance equals 1, such that, for $g > 1$, at $L = L_*$ 
the device crosses
over from  the weakly to the strongly coupled regime.

\subsection{The weakly coupled fixed point}

At the weakly coupled fixed point ($E_J \to 0$), the field $\Phi ( x
, t)$ obeys Dirichlet boundary conditions at the inner boundary
and Neumann boundary conditions at the outer boundary. As a consequence, 
$\Phi ( x , t )$ takes the mode expansion

\beq 
\Phi ( x , t ) =  - \sqrt{\frac{1}{g}} \sum_{ n \neq 0} \frac{
\alpha ( n )}{ n} \sin ( k_n x ) e^{ - i k_n u t } 
\:\:\:\: ,
\label{for8bis} 
\eneq 
\noindent 
with $[ \alpha ( n ) , \alpha (n' )
] = n \delta_{ n + n' , 0 }$, and $k_n = \frac{  \pi}{L} \left( n +
\frac{1}{2} \right)$. 

The ground state of the bosonic theory, $| {\rm GS} \rangle$, is defined
in terms of the oscillatory modes as

\beq
\alpha ( n ) | {\rm GS} \rangle = 0 \:\:\: ; \:\: \forall  > 0 
\:\:\:\: . 
\label{new1}
\eneq
\noindent

Normal ordering the Josephson energy with respect to $|{\rm GS} \rangle$, 
leads to

\beq
H_J = - \frac{E_J}{2}  \left( \frac{2 \pi  a}{L} \right)^\frac{1}{g}
\left[ : e^{ i \sqrt{2} \Phi ( L ) + i \varphi} : + 
 : e^{ - i \sqrt{2} \Phi ( L ) - i \varphi} : \right]
\:\:\:\: .
\label{for12}
\eneq
\noindent
where the column $::$ denotes, as usual,  normal ordering.

The thermodynamic limit (i.e., large $L$) is attained once the
scaling equations for the pertinent running couplings have been
obtained using the renormalization group (RG) approach (see
appendix A).

The RG equations are determined by the requirement that the
partition function ${\bf Z}$ is independent of the cutoff. In order to
compute ${\bf Z}$, one should Wick-rotate the field $\Phi$ and resort to
the imaginary-time Feynman path-integral formalism. The partition
function is, then, given by

\beq {\bf Z} = \int {\bf D} \Phi e^{ - S_0 } {\bf T}_\tau [ e^{- S_B} ]
\:\:\:\: , \label{eqaddi1.1.3} \eneq \noindent where ${\bf T}_\tau$
is the imaginary time ordering operator, $S_0$ is the free Euclidean
action

\beq
S_0 = \frac{g}{2 \pi}  \int_0^\beta \: d \tau \: \int_0^L \: d x \:
\left[ \frac{1}{\vpl} \left( \frac{ \partial \Phi}{ \partial \tau } \right)^2
+ \vpl \left( \frac{ \partial \Phi}{ \partial x} \right)^2 \right]
\:\:\:\: ,
 \label{eqaddi1.1.4}
\eneq \noindent and $S_B$ is the boundary interaction action at
imaginary times

\beq
S_B = - \frac{ \bar{E}_J }{2} \int_0^\beta \: d \tau \:
\left[ : e^{ i [ \sqrt{2} \Phi ( i \tau ) + \varphi ]}: 
+  : e^{ - i [ \sqrt{2} \Phi ( i \tau ) + \varphi ]}: \right]
 \:\:\:\: ,
\label{eqaddi1.6} \eneq \noindent 
with $\Phi ( i \tau ) = \Phi ( L , i \tau )$.

In Eq.(\ref{eqaddi1.6}), the
renormalized coupling constant $\bar{E}_J = E_J ( 2 \pi a /
L)^\frac{1}{g}$ has been introduced. Of course, after this
redefinition of the Josephson coupling, there is not anymore a
cutoff dependence  of the interaction operator \cite{coleman}.

The boundary action in Eq.(\ref{eqaddi1.6}) contains only the vertex
operators $V_{  \pm 1  } ( \tau ) =
 e^{ \pm i \varphi} : \exp \left[ \pm i \sqrt{2} \Phi ( i \tau )\right] :$. 
Due to the mixed boundary conditions on $\Phi$, the
operator $\Phi ( i \tau )$ contains no zero modes and, thus, there
are no selection rules on the Kac-Moody charge carried by the vertex
operators. As a consequence, the Operator Product Expansions
(O.P.E.'s) may generate additional vertices, with higher periodicity
in $\varphi$, given by

\beq
V_{ n }  ( \tau )
 =  e^{ \pm n i \varphi} : \exp \left[ \pm i \sqrt{2} n \Phi ( i \tau ) \right]
: \:\:\:\: .
\label{e3.4}
\eneq
\noindent

The scaling dimension of $V_{ n }$, $h_n$, is derived from the
correlation function $ \langle T_\tau [ V_{ n } ( \tau )  V_{  - n }
( \tau )  ] \rangle$ which, for $\beta \to \infty$, is approximatively given
by (see Appendix B)

\beq \langle T_\tau [ V_{ n } ( \tau )  V_{  - n } ( \tau' )  ] \rangle
\approx  \left| \tanh \left[ \frac{ \pi \vpl }{4 L } ( \tau - \tau'  ) \right] 
\right|^{ - \frac{ 2 n^2}{g} } 
 \;\;\;\; . \label{e3.5.0} \eneq
\noindent

At short imaginary time distances ($ \vpl | \tau - \tau' | / L \ll 1$) , 
Eq.(\ref{e3.5.0}) becomes

\beq
 \langle T_\tau [ V_{ n } ( \tau )  V_{  - n } ( \tau' )  ] 
\rangle \approx  \left|  \frac{  \pi \vpl }{L} | \tau - \tau' |
\right|^{ - \frac{ 2 n^2}{g} } \;\;\;\;, \label{e3.5} \eneq
\noindent

leading to   $ h_n =   n^2 / g$.

As we shall see in the following, despite the fact that $h_n$ depends on $n$,
due to nonlinearities, the higher harmonics in Eq.(\ref{e3.4})
are irrelevant operators, when $g < 1$, while they become relevant for $g > 1$.

The O.P.E. yields the pertinent fusion rules for the vertex operators

\beq V_n ( \tau ) V_{ n' } ( \tau' ) \approx_{ \tau' \to \tau }
\left[ \frac{  \pi \vpl}{L} ( \tau - \tau' ) \right]^{ - h_n - h_{n'}
+ h_{ n + n'} } V_{ n + n' } ( \tau ' ) \;\;\;\; . \label{e3.6}
\eneq \noindent

Introducing the pertinent counterterms, one may easily show  that the effective
boundary action looks as

\beq
S_{B , Eff} = - \frac{1}{2}  \int_0^\beta \: d \tau \: \sum_{ n
= 1}^\infty  \biggl\{ \bar{E}_n e^{ i n  \varphi } V_n ( \tau ) +
\bar{E}_n e^{ - i n  \varphi } V_{ - n } ( \tau ) \biggr\}
\:\:\:\: ,
\label{e3.7}
\eneq
\noindent

where the renormalization condition is that all the coupling
strengths $\bar{E}_n = ( 2 \pi a / L )^{\frac{ n^2}{g}} E_n$,
 with $n \neq \pm 1$, vanish at a given reference
length scale $L_0$.

Setting the running couplings $g_n$ as

\beq
g_n = \left( \frac{L}{2 \pi a}\right)^{1 - \frac{  n^2 }{g }} \: \frac{a^n
E_n}{\vpl^n}
 \:\:\:\: ,
\label{e3.8}
\eneq
\noindent

from the  fusion rules given in Eq.(\ref{e3.6}), one is  able
to derive, in principle,  the $\beta$-function for all the couplings $g_n$.

For the sake of simplicity, we shall restrict our attention to the
first two relevant terms, proportional to the  running couplings
$g_1,g_2$. The RG equations are, then

\beq \frac{ d g_1}{ d \ln ( L / L_0 )} = \beta_1 ( g_1 , g_2 ) =
\left( 1 - \frac{1}{g} \right) g_1 + g_1 g_2 \:\:\:\: , \label{e3.9}
\eneq \noindent and

\beq
\frac{ d g_2}{ d \ln ( a / a_0 ) } = \beta_2 ( g_1 , g_2 ) =
 \left( 1 - \frac{4}{g} \right) g_2
+ ( g_1 )^2
\:\:\:\: :
\label{e3.10}
\eneq
\noindent

the flux $\varphi$ fully decouples from
Eqs.(\ref{e3.9},\ref{e3.10}) and, thus, the scaling near the Neumann fixed
point is independent of $\varphi$.

For $g < 1$, $g_1$ and $g_2$ are both
irrelevant. Thus, the Neumann fixed point is stable and the theory
is perturbative in the boundary interaction with effective,
size-dependent couplings given by

\beq
g_1 ( L ) \approx g_1 (L_0 ) \left( \frac{L}{L_0} \right)^{
\left( 1 - \frac{1}{g} \right) } \;\;\; ; \;\; g_2 ( L ) \approx 0
\:\:\:\: .
\label{e3.11b}
\eneq
\noindent

 On the other hand, when $g > 1$, $g_1$ becomes relevant since it  grows
with $L$ as

\beq g_1 ( L ) \approx g_1 ( L_0 ) \left( \frac{L}{L_0} \right)^{
\left( 1 - \frac{1}{g} \right)} \:\:\:\: . \label{ecco} \eneq
\noindent By inserting Eq.(\ref{ecco}) into the nonlinear term of
Eq.(\ref{e3.10}), one gets

\beq
g_2 ( L ) \approx \frac{ ( g_1^2 (L ) )}{ 1 + 2 / g }
 \left[ 1 - \left( \frac{L}{L_0} \right)^{ -1 - \frac{2}{g}} \right]
 \:\:\:\: ,
\label{e3.15} \eneq \noindent where the integration constants have
been chosen so as to be consistent  with the renormalization
condition $g_2 ( L_0 ) = 0$.

From Eqs.(\ref{e3.15}), one sees that the nonlinear term in
Eq.(\ref{e3.9}) makes $g_2 (L)$ increase, as soon as one starts the
scaling flow, even when $ 1 - 4 / g < 0$. This is a remarkable
result, since it implies
 that,  as soon as  the first  harmonics in $\varphi$ sets in,
all the others follow, due to nonlinearities in the RG equations.

At the Neumann fixed point,  the Josephson current is perturbative in
$\bar{E}_J$: of course, as $L$ increases, our results are
reliable only for $g < 1$. As a function of the flux
$\varphi$, $I [ \varphi ]$, is given by

\beq
 I [ \varphi ] =  \frac{ 1}{c  \Phi_0^* } \lim_{ \beta \to \infty} 
\frac{1}{ \beta } \frac{ \partial }{\partial \varphi } \ln \left[  
\frac{  {\bf Z}_{Eff} [ \varphi ] }{ {\bf Z}_0 } \right]
\:\:\:\: ,
\label{eqaddi1.10}
\eneq
\noindent

where  ${\bf Z}_{Eff} [ \varphi ]$ is the $\varphi$-dependent 
partition function, given by

\beq 
{\bf Z}_{Eff} [ \varphi ] = 
\int {\bf D} \Phi e^{ - S_0 } {\bf T}_\tau [ e^{- {S}_{B , Eff}} ]
\:\:\:\: .
\label{eqaddi1.9}
\eneq
\noindent

To the second order in $g_n$, one gets

\beq
I [ \varphi ] =  e^* E_J \left( \frac{ 2 \pi a}{L}
\right)^\frac{1}{g} \left[ \sin ( \varphi ) + \frac{  g_2 ( L ) }{ g_1 ( L ) }
\sin ( 2 \varphi ) \right]
\:\:\:\: ,
\label{josc}
\eneq
\noindent
which explicitly gives the contribution of the second harmonics to the
Josephson current.

For $g>1$, Eq.(\ref{e3.15}) shows that, for $L \to \infty$,
the ratio between the running coupling constants of the first two
harmonics is given by

\beq \frac{g_2 ( L) }{g_1 ( L ) } \propto \left( \frac{L}{L_0}
\right)^{1 - \frac{1}{g}} \:\:\:\: , \label{nuo1} \eneq \noindent
showing the relevance of the boundary interaction, since  higher
harmonics become more and more relevant, as $L \to \infty$. 
The scaling equations (\ref{e3.8},\ref{e3.9},\ref{e3.10},\ref{e3.11b})
cease to be valid  at the renormalized healing length $\bar{L}_* $ for which 
$ g_1 ( \bar{L}_* ) \sim 1$, i.e., when 

\beq
L = \bar{L}_* = 2 \pi  a \left( \frac{\vpl}{a E_J} \right)^\frac{g}{g-1}
 \:\:\:\: ,
\label{nuo2}
\eneq
\noindent

yielding that the renormalized Josephson coupling 
$ \bar{E}_{J}^* = \bar{E}_{J} ( L_* )$ is given by

\beq
\bar{E}_{J}^* = \left( \frac{a E_J}{\vpl} \right)^\frac{1}{g - 1} E_J
 \:\:\:\:, 
 \eneq
\noindent

which is the energy where the level spacing of the plasmon field
 becomes of the order of the Josephson energy, i.e., $ 2 \pi \vpl / \bar{L}_*
\sim ( 2 \pi a / \bar{L}_* )^\frac{1}{g} E_J$ \cite{glazhek}.

\subsection{The strongly coupled fixed point}

In this subsection, we analyze the properties of the rf-SQUID in the
limit in which $\bar{E}_J / ( 2 \pi \vpl /L )$ is $\gg 1$: $H_J$ is
now the dominating potential term, and the field $\Phi ( L , t )$
takes values corresponding to a minimum of the boundary energy.
Therefore, at both boundaries $\Phi (x  )$ must obey to Dirichlet
boundary conditions, given by

\beq
\Phi ( 0 , t) = 0 \:\:\: ; \:\: \sqrt{2} \Phi ( L , t ) = - \varphi  + 2 \pi k
 \:\:\: ; \:\: (k \in Z)
\:\:\:\: .
\label{nuo4.1}
\eneq
\noindent

Both boundary conditions are satisfied if one chooses for $\Phi ( x , t )$
the mode expansion

\beq
\Phi ( x , t ) = - \frac{2 \pi x}{L}  P -
 \sqrt{\frac{1}{g}} \sum_{ n \neq 0} \frac{\alpha (n )}{n}
\sin ( k_n x ) e^{ i k_n \vpl t } 
\:\:\:\: ,
\label{nuo4.2}
\eneq
\noindent
where $P$ is the zero mode operator and
$k_n = \left( \frac{ \pi}{L} n \right)$.  From Eq.(\ref{nuo4.1}), one obtains
the set of possible eigenvalues of $P$:

\beq
P_k = \frac{1}{\sqrt{2}} \left[ - \frac{\varphi}{2 \pi}  +  k \right]
\:\:\:\: .
\label{nuo4.2b}
\eneq
\noindent
As a consequence, there will be an infinite set of ground states of the
oscillatory modes, corresponding to the possible eigenvalues of $P$, and
denoted by $ | P_k , \{ 0 \} \rangle$.

Once $H_{rf}$ has been evaluated for the particular solution given in
Eq.(\ref{nuo4.2}), one obtains  the effective Hamiltonian at the
Dirichlet fixed point, $H_D$,

\beq H_D = \frac{  \pi \vpl }{ L} \left[  2 g P^2   + 2  \sum_{ n =
1}^\infty  \alpha ( - n ) \alpha ( n ) \right]
\:\:\:\: ,
\label{nuo4.5} \eneq \noindent

from which one gets that the partition function, 
at the strongly coupled fixed point

\beq {\bf Z}_D [ \varphi ] = {\rm Tr} [ e^{ - \beta H_D} ] = \frac{1}{
\eta ( q ) } \sum_{ k \in Z} \exp \left[ - \beta \frac{ g \pi \vpl}{
 L}  \left( - \frac{\varphi}{2 \pi}   +   k \right)^2 \right]
\:\:\:\: , \label{nuo4.6} \eneq \noindent with $ \eta ( x ) =
\prod_{ n = 1}^\infty ( 1 - x^n )$, and $q = \exp \left[ - \beta
\frac{  \pi  \vpl}{L} \right]$,

leading to the well-known sawtooth-like dependence on $\varphi$ of 
the Josephson current as $\beta \to \infty$ \cite{glazhek}, since

\beq I [ \varphi ] = - \lim_{\beta \to \infty} \frac{1}{\beta}
\frac{ \partial \ln {\bf Z}_D [ \varphi ] }{
\partial \varphi } \propto \varphi - [ \varphi ]
\:\:\:\: ;
 \label{nuo4.7}
\eneq
\noindent

of course, $ [ \varphi ]$ is the integer part of $\varphi$ (in units
of $ 2 \pi$).

At strong coupling, the degeneracy among the minima of the boundary
term is removed by the magnetic energy, $E^{(0)}_k = \frac{ g \pi  \vpl}{  L}
 \left( - \frac{\varphi}{2 \pi}   +   k \right)^2$, selecting
only one eigenvalue of the zero mode of the field $\Phi ( x , t )$, except 
when $\varphi = \pi + 2 \pi k$, where $E^{(0)}_k = E^{(0)}_{k + 1}$.
When these two levels are confined away from the others, 
 the rf-SQUID may operate as a qubit \cite{makhlin}.
For $L > L_*$, however, the oscillatory modes of the plasmon field determine a
renormalization of the parameters of the effective two-level system,
as it will be discussed  in the following.

\subsection{Renormalization of the Josephson current for the rf-SQUIDs}

At strong coupling and for $\varphi = \pi$, the ground state is
twofold degenerate, that is, $E^{(0)}_0 = E^{(0)}_{ 1}$. This degeneracy
is removed by phase slips (instantons), of
amplitude $\sqrt{2} \pi $, connecting the ground states.

In order to describe  phase slips within the framework of the two-boundary
sine Gordon model, one has to introduce the dual field of $\Phi ( x , t )$, 
$\Theta ( x , t )$, defined by

\beq
\frac{ \partial \Phi ( x , t ) }{ \partial x} = \frac{\vpl}{2 g} \frac{ 
\partial \Theta ( x ,  t ) }{ \partial t} 
\:\:\: ; \;\; 
\frac{2g }{\vpl} 
\frac{ \partial \Phi ( x , t ) }{ \partial t} = \frac{ 
\partial \Theta ( x ,  t ) }{ \partial x} 
\:\:\:\: . 
\label{dual2}
\eneq
\noindent

Due to the duality between $\Phi$ and $\Theta$, the effective Hamiltonian 
given in Eq.(\ref{nuo4.5}) may be usefully rewritten as

\beq
H_D = \frac{1}{ 4 \pi (2g) } \: \int_0^L \: d x \: 
\left[ \frac{1}{\vpl} \left( \frac{ \partial \Theta }{ \partial t} \right)^2 + 
\vpl \left( \frac{ \partial \Theta }{ \partial x} \right)^2 \right]
\:\:\:\: . 
\label{ade34}
\eneq
\noindent

From the  algebra of the bosonic fields,
one easily derives \cite{wires} the commutation relations

\beq
\left[ \frac{ \partial \Phi ( x , t ) }{ \partial x } , \Theta ( y , t ) 
\right] = 
\left[ \frac{ \partial \Theta ( x , t ) }{ \partial x } , \Phi ( y , t ) 
\right]  = 2 \pi i \delta ( x - y ) 
\:\:\:\: . 
\label{dual3}
\eneq
\noindent

In order to construct the leading correction to the Dirichlet fixed 
point, one may use the Delayed Evaluation of Boundary Conditions (DEBC)
approach, introduced in Ref.\cite{wires}: for this purpose,  one has first to
consider the most general primary field of the bulk theory, namely,
a vertex operator of the form $V_{n , \tilde{n}} ( x , t ) = 
: \exp \left[ i \frac{n}{ \sqrt{2}}  \Phi ( x , t ) + i \frac{ \tilde{n}}{ 
\sqrt{2}}  \Theta ( x , t ) \right] :$, and then to obtain the boundary 
perturbation, $V^{(B)}_{ n , \tilde{n}} ( t )$,  by evaluating it at  
$x = L$, using  the pertinent boundary 
conditions for the fields $\Phi ( x , t )$ and $\Theta ( x , t )$. Since 
 $\Phi ( x , t )$ obeys to the Dirichlet boundary conditions 
 at $x=0$ and at $x=L$, it will
not contribute to   $V^{(B)}_{ n , \tilde{n}} ( t )$. At variance, 
 $\Theta ( x , t )$, obeys to the Neumann boundary
conditions at both boundaries;  namely, one has that 

\beq
\frac{ \partial \Theta ( 0 , t )}{ \partial x} = 
\frac{ \partial \Theta ( L , t )}{ \partial x}
= 0 
\:\:\:\: 
\label{dual4}
\eneq
\noindent

implying the following mode expansion for $\Theta ( x , t )$

\beq
\Theta ( x , t ) =  \theta_0 + \frac{ 2 \pi u t  }{L} 2 g P   
+ 2 i \sqrt{ g } \sum_{ n \neq 0} \frac{ \alpha ( n )}{n} \cos \left[ 
\frac{ \pi n }{L}  x \right] 
e^{ - i k_n t } 
\:\:\:\: , 
\label{dual5}
\eneq
\noindent
with $\theta_0 = ( q_R - q_L ) / \sqrt{2} $ and $k_n = \pi n / L$. 

As a consequence of Eq.(\ref{dual5}), one finds

\beq
\Theta ( i \tau ) =  \Theta ( L , i \tau ) = 
 \theta_0 + \frac{ 2 \pi i u \tau }{L} 2 g P 
+ i 2 \sqrt{ g } \sum_{ n \neq 0} \frac{ \alpha ( n )}{n} (-1)^n 
e^{ -  \frac{ \pi n}{L} \tau } 
\:\:\:\: 
\label{dual6}
\eneq
\noindent

and thus, the generic boundary perturbation at the strongly coupled fixed 
point, $\tilde{H}_{\tilde{B}}$, may be written as a linear combination of the 
vertices $V_{\tilde{n}} =  V^{(B)}_{0 , \tilde{n}}$ as

\beq
\tilde{H}_{\tilde{B}} 
= - \sum_{ \tilde{n}} \lambda_{ \tilde{n}} : e^{ i \frac{ \tilde{n} }{ 
\sqrt{2}} \Theta ( L ) }:
\:\:\:\: . 
\label{dual7}
\eneq
\noindent

At the Dirichlet fixed point, the partition function  
$\tilde{Z}_D [ \varphi ]$, including
the boundary interaction in Eq.(\ref{dual7}),
is written as

\beq 
\tilde{\bf Z}_D [ \varphi ] = {\rm Tr} \left[ e^{ - \beta H_D}  {\bf T}_\tau 
\exp \left\{ \sum_{ \tilde{n}  \neq 0  } \int_0^\beta 
\: d \tau \: \lambda_{ \tilde{n}} 
: e^{ i \frac{ \tilde{n} }{ \sqrt{2} }  \Theta ( i \tau ) }: \right\}  \right]
\:\:\:\: ,
\label{dual8}
\eneq
\noindent
with $\Theta ( i \tau ) = \Theta ( L , i \tau)$. 

Eq.(\ref{dual8}) reduces to Eq.(\ref{nuo4.6}) when all the $\lambda_n$'s 
are set to zero.

Following the same steps used  in section (2.1), one may derive the scaling 
dimension of   $V_{\tilde{n}}$, $h_{ \tilde{n}}$,
 from the large-$L$-limit of the two point correlation
function; namely, 

\beq
\langle {\bf T}_\tau [ V_{\tilde{n} } ( i \tau )  V_{ - \tilde{n} } 
( i \tau' )  ] \rangle_{(0)} \approx
\left|  \frac{  \pi \vpl }{L} | \tau - \tau' |
\right|^{ -  2 g \tilde{n}^2  } \;\;\;\;, \label{dual9} \eneq
\noindent

where $\langle \ldots \rangle_{(0)}$ denotes thermal averaging defined by
the free Hamiltonian with Dirichlet boundary conditions for $\Phi ( x , t )$
at both boundaries.

Eq.(\ref{dual9}) implies that $h_{ \tilde{n}} = g  \tilde{n}^2 $ and that
 the leading perturbation at the strongly coupled fixed point is given by

\beq
H_{\tilde{B}} = - Y [ [ : e^{ i  
\frac{ \Theta (L)}{ \sqrt{2}}  } : + : e^{ - i  \frac{ \Theta (L)}{ \sqrt{2}} 
} :  ]
\:\:\:\: . 
\label{dual10}
\eneq
\noindent

The physical meaning of the operators $ V_{\tilde{n} }$ may be inferred
from the commutation relation

\beq
[ P ,: e^{ i \tilde{n} \Theta ( i \tau )   } : ] =
 - \frac{1}{2 \pi }     \int_0^L \: d x \: 
\left[ \frac{ \partial \Phi ( x , i \tau ) }{ \partial x} , 
: e^{ i \frac{ \tilde{n} }{ \sqrt{2}}  \Theta ( t )   } : \right] = 
\frac{  \tilde{n} }{ \sqrt{2}} : e^{ i \frac{ \tilde{n} }{ \sqrt{2}}
 \Theta ( i \tau ) } :
\:\:\:\: ,
\label{dual11}
\eneq
\noindent
which shows that  $ V_{\tilde{n} }$ changes the 
eigenvalue of $P$ by $  \tilde{n} / \sqrt{2}$. Thus, the boundary
fields  $: e^{ \pm \frac{ i}{\sqrt{2}}  \Theta ( i \tau) } :$ describe
instanton/antiinstanton trajectories between the two ground states that
are degenerate at $\varphi = \pi$ and the parameter 
$Y$ may be interpreted as the instanton fugacity. 

A dimensionless running coupling $Y_{\rm Inst} ( L)$ is defined as

\beq
Y_{\rm Inst} ( L ) = Y L^{ 1 - g } 
\:\:\:\: . 
\label{dual12}
\eneq
\noindent
From Eq.(\ref{dual12}),  one
sees that, for $g<1$,  (i.e., when the Neumann fixed point is
stable), instantons are a relevant perturbation,
since $Y_{\rm Inst} ( L)$  scales as $\left( \frac{L}{L_* }
\right)^{1-g}$. However, the scaling ceases to be valid 
 as  $Y_{\rm Inst} ( L ) \sim 1$.

For $g>1$,  using the boundary perturbation operator introduced 
in Eq.(\ref{dual10}),
it is straightforward  to compute the leading corrections to the Josephson
current. For instance,  when  $\varphi \sim 0$, one may approximate the
partition function in Eq.(\ref{dual8}) as

\[
\tilde{\bf Z}_D 
[ \varphi ] \approx \frac{ \exp \left[ - \beta \frac{ g  \vpl }{
4 \pi L } \varphi^2 \right] }{ \prod_{ n > 0 } 
[ 1 - q^n ] } \langle P_0 , \{ 0 \} | \biggl\{ 1 + 
\]

\beq
2 Y^2 \int_0^\beta \: d \tau_1 \: d \tau_2 \: {\bf T}_\tau [ : \cos (  
\frac{1}{ \sqrt{2}} \Theta ( i \tau_1 )) : 
: \cos ( \frac{1}{ \sqrt{2}} \Theta ( i \tau_2 ) ) : ] \biggr\} 
| P_0 , \{ 0 \} \rangle
\:\:\:\: . 
\label{ade40}
\eneq
\noindent

By inserting in Eq.(\ref{ade40}) the explicit form of the correlator
\beq
 \langle P_0 , \{ 0 \} |  {\bf T}_\tau [ : \cos ( \frac{ \Theta ( 
i \tau_1 )  }{ \sqrt{2}} )  : : \cos ( \frac{ \Theta ( i \tau_2 ) }{ \sqrt{2}} 
) : ]  | P_0 , \{ 0 \} \rangle =  \frac{ \cosh \left[ \frac{ g  \vpl 
\varphi }{2 L} ( \tau_1 - \tau_2 ) \right]}{
\left[ 2 \sinh \left[ \frac{ \pi \vpl }{2 L} | \tau_1 - \tau_2 | 
\right] \right]^{2g} }
\:\:\:\: ,
\label{ade41}
\eneq
\noindent

one gets the partition function ${\bf Z} [ \varphi ]$

\beq
\tilde{\bf Z}_D 
[ \varphi ] \approx \frac{ \exp \left[ - \beta \frac{g \vpl}{4 \pi L } 
\varphi^2 \right] }{ \prod_{ n > 0 }  [ 1 - q^n ] }
\biggl\{ 1 + 
\frac{ \beta L}{ \pi \vpl } Y^2 
\Gamma [ 1 - 2 g ] \biggl\{  \frac{ \Gamma [ g  ( 1 - 
 \frac{ \varphi}{2 \pi} )  ] }{
\Gamma [ 1 -  g ( 1 +  \frac{ \varphi}{2 \pi} )  ] } +
\frac{ \Gamma [  g ( 1 + 
 \frac{ \varphi}{2 \pi} )  ] }{
\Gamma [ 1 -  g ( 1 -  \frac{ \varphi}{2 \pi} )  ] } \biggr\}
\biggr\}
\:\:\:\: . 
\label{ade43}
\eneq
\noindent

Computing the logarithmic derivative of Eq.(\ref{ade43}) with respect to 
$\varphi$ and dividing it by $- \beta$, one obtains

\[
I [ \varphi ] \approx \frac{ e^* g \vpl }{2 \pi L } \varphi - 
e^*  \frac{  2 L \Gamma [ 1 - 2 g ] }{ 2 \pi \vpl } Y^2  
\frac{ \partial }{\partial \varphi} \biggl\{ \frac{ \Gamma [ g  
( 1 -  \frac{ \varphi}{2 \pi} )  ] }{
\Gamma [ 1 -  g ( 1 +  \frac{ \varphi}{2 \pi} )  ] } +
\frac{ \Gamma [  g ( 1 + 
 \frac{ \varphi}{2 \pi} )  ] }{
\Gamma [ 1 -  g ( 1 -  \frac{ \varphi}{2 \pi} )  ] } \biggr\}
\]

\beq
\approx  \left[ \frac{ e^* g \vpl }{2 \pi L } - Y^2   G [ g ] \right] \varphi
\:\:\:\: , 
\label{ade44}
\eneq
\noindent

with

\beq
G [ g ] = e^* \frac{ 2 L \Gamma [ 1 - 2 g ] }{ 4 \pi^2 \vpl} 
\frac{ \partial^2 }{\partial \varphi^2} \biggl\{ \frac{ \Gamma [ g  
( 1 -  \frac{ \varphi}{2 \pi} )  ] }{
\Gamma [ 1 -  g  ( 1 +  \frac{ \varphi}{2 \pi} )  ] } +
\frac{ \Gamma [  g ( 1 + 
 \frac{ \varphi}{2 \pi} )  ] }{
\Gamma [ 1 -  g ( 1 -  \frac{ \varphi}{2 \pi} )  ] } \biggr\}_{
\varphi = 0 }
\:\:\:\: . 
\label{ade44bis}
\eneq 
\noindent

For small $\varphi$, $Y$, $H_{\tilde{B}}$ provides a small correction 
to the slope of the Josephson current. As we shall see in detail in the
following subsection, for $\varphi = \pi$ and  $g > 1$,
instantons at the Dirichlet
fixed point  remove the degeneracy between the minima and
 smooth the edges in the sawtooth-like Josephson current.
For $g<1$, of course, instantons are irrelevant operators. 

\subsection{Instanton constribution to the Josephson current at the Dirichlet
fixed point for $\varphi = \pi $.}

In the previous section, it has been shown that, for  $\varphi = \pi$, 
the state $P_0 , \{ 0 \} \rangle$  becomes degenerate with  $P_1 , \{ 0 \} 
\rangle$.  This corresponds to a discontinuous jump of $e^* g \vpl / L$ in
the Josephson current, between the values corresponding  to the states that
are degenerate at $\varphi = \pi$. Since instantons 
 connecting the two states should remove  the degeneracy, for $Y \neq 0$
one should expect that the Josephson current becomes a continuous function of 
$\varphi$ at $\varphi = \pi$.

In order to compute the $Y$-dependent corrections to the ground state
energy, it is useful to partition the Hilbert space into subspaces
on which  $H_{\tilde{B}}$ is diagonal.  If 
$ | P_k , \Psi \rangle$ denotes a generic state in the sector corresponding 
to the eigenvalue $P_k$ of the zero-mode operator, one defines the states
belonging to the above subspaces as 

\[
| \pm , \Psi \rangle = \frac{1}{ \sqrt{2}} [ | P_0 , \Psi \rangle
\pm | P_1 , \Psi \rangle ] 
\:\:\:\: .
\]
\noindent
Thus,  the partition function at $\varphi = \pi$
 may be approximated as

\beq
\tilde{\bf Z}_D [ \pi ]  =  {\rm Tr}_{ | - , \Psi \rangle } 
\biggl\{ e^{ - \beta H_D } 
{\bf T}_\tau
\exp [ \int_0^\beta \: d \tau \: \tilde{H}_B ( i \tau ) ]  \biggr\}
+ {\rm Tr}_{ | + , \Psi \rangle } 
\biggl\{ e^{ - \beta H_D }{\bf T}_\tau
\exp [ \int_0^\beta \: d \tau \: \tilde{H}_B ( i \tau ) ]  \biggr\}
\:\:\:\: , 
\label{ade51}
\eneq
\noindent
where ${\rm Tr}_{ | \mp , \Psi \rangle }$ 
denotes tracing over the sector of the Fock
space spanned by the states $ | \pm , \Psi \rangle $.

At low temperature, Eq.(\ref{ade51}) may be approximated as

\[
\tilde{\bf Z}_D [ \pi ] \approx \langle - , \{ 0 \}   |  e^{ - \beta H_D } 
{\bf T}_\tau \exp [ \int_0^\beta \: d \tau \: \tilde{H}_B ( i \tau ) ]  
| - , \{ 0 \}   \rangle
\]

\beq
+ \langle + , \{ 0 \}  |  e^{ - \beta H_D } 
{\bf T}_\tau \exp [ \int_0^\beta \: d \tau \: \tilde{H}_B ( i \tau ) ]  
| + ,  \{ 0 \} \rangle
\:\:\:\: ,
\label{ade52}
\eneq
\noindent

where $| \pm , 0 \rangle $ are, of course,  the ground states of both
subspaces.  
By expanding ${\bf T}_\tau \exp [ \int_0^\beta \: d \tau \: 
\tilde{H}_B ( i \tau ) ]  $ in a power series of $Y$, one gets

\[
\langle \pm ,  \{ 0 \} |  e^{ - \beta H_D } 
{\bf T}_\tau \exp [ \int_0^\beta \: d \tau \: \tilde{H}_B ( i \tau ) ]  
| \pm ,  \{ 0 \}  \rangle
= 
\]

\beq
\exp \left[ - \beta \frac{ g \pi \vpl }{ 4 L }  \right]
\sum_{ m = 0 }^\infty  \frac{1}{m !} \int_0^\beta \: d \tau_1  \: \ldots \:
 \int_0^\beta \: d \tau_m  \langle \pm , \{ 0 \} | {\bf T}_\tau [
  \tilde{H}_B ( i \tau_1 ) \ldots  \tilde{H}_B ( i \tau_m ) ] 
| \pm ,  \{ 0 \} \rangle
\:\:\:\: . 
\label{ade53}
\eneq
\noindent

When computing the partition function, one has to sum over  contributions
from connected, as well as from disconnected diagrams. By applying
Wick's theorem, one obtains

\[
 \langle \pm , \{ 0 \} |  {\bf T}_\tau [ \tilde{H}_B ( i \tau_1 )
\ldots  \tilde{H}_B ( i \tau_m ) ] | \pm ,  \{ 0 \} \rangle
= \prod_{j = 1}^M  \langle \pm , \{ 0 \}|  \tilde{H}_B ( i \tau_1 ) 
| \pm ,  \{ 0 \}  \rangle + 
\]

\beq
\sum_{ j_1 \neq j_2 = 1}^M 
 \langle \pm ,  \{ 0 \} |   {\bf T}_\tau [ \tilde{H}_B ( i \tau_{j_1}  ) 
\tilde{H}_B  ( i \tau_{j_2}  )  ] | \pm ,  \{ 0 \}  \rangle 
\prod_{ i \neq  j_1 , j_2 }  \langle \pm , \{ 0 \} |
\tilde{H}_B ( i \tau_i )  | \pm ,  \{ 0 \}  \rangle + \ldots 
\:\:\:\: . 
\label{ade54}
\eneq
\noindent

Since $\langle \pm , {\rm GS} |  \tilde{H}_B ( i  \tau ) 
| \pm , {\rm GS} \rangle = \pm Y$, 
it is straightforward to evaluate, using Eqs.(\ref{ade53}, \ref{ade54}),
the leading $Y$-dependent contribution to the energy of the states 
$ \pm , \{ 0 \} \rangle$; namely $E_\pm$ is given by

\beq
E_\pm =  \frac{ g \pi \vpl }{ 4 L } \mp Y - \frac{ L}{\pi \vpl} Y^2 
\Gamma [ 1 - 2 g ] \biggl\{ \frac{ \Gamma [ \frac{g}{2}  ]}{ 
\Gamma [ 1 - \frac{3}{2} g  ]} +  \frac{ 
\Gamma [ \frac{3}{2} g  ]}{ \Gamma [ 1 - \frac{g}{2}  ]}
\biggr\}
\:\:\:\: ,
\label{ade56}
\eneq
\noindent

and, thus, even to the first order in $Y$, instantons remove the
degeneracy between the ground states.

The observable consequence of the removal of the ground state degeneracy
 is the smoothing down of the Josephson current, which 
becomes a continuous function of the applied flux $\varphi$ when it
crosses the value $\varphi  = \pi$. 

In order to see this effect, one has to resum over instanton contributions
to the partition function, by setting  $\varphi = \pi + \epsilon$, 
with $\epsilon / \pi \ll 1$. In order to obtain a meaningful result,
one has first  to sum over the
instanton contributions, and then to compute the leading corrections in
$\epsilon$.

When  $\varphi = \pi + \epsilon$, the low temperature partition function 
 may be approximated as 

\beq
\tilde{\bf Z}_D [ \pi + \epsilon ]  =  \sum_{ a = \pm } 
{\rm Tr}_{ | a , \Psi \rangle } 
\biggl\{ e^{ - \beta H_D } 
{\bf T}_\tau
\exp [ - \delta S [ Y , \epsilon ] ] \biggr\}
\:\:\:\: , 
\label{ade51.1}
\eneq
\noindent

with

\beq
\delta S [ Y , \epsilon ] = 2 Y \int_0^\beta \: d \tau \: 
\cos \left[ \frac{ \Theta ( i \tau ) }{ \sqrt{2}} \right] +
\beta \left[ \frac{g \vpl}{ L } \epsilon P + \frac{ g 
\epsilon^2 }{ 4 \pi L } \right]
\:\:\:\: . 
\label{ade57}
\eneq
\noindent 
When $\epsilon = 0 $, one obtains the result of the previous section. When
computing the leading corrections in $\epsilon$, one has to notice that
$P$ mixes the two subspaces labeled by  $\pm$; namely

\beq
\langle \pm , \Psi | P | \mp ,  \Psi \rangle = \mp \frac{1}{ 2 \sqrt{2}}
\:\:\:\: . 
\label{ade60}
\eneq
\noindent

Thus, the leading $\epsilon$-dependent contribution to Eq.(\ref{ade57}),
$\delta \tilde{\bf Z}_D [ \epsilon ]$, appears only to the 
order $\epsilon^2$, and is given by

\[
\delta \tilde{\bf Z}_D [ \epsilon ] = 
\tilde{\bf Z}_D [ \pi ] e^{ - \beta \frac{ g \vpl }{ 4
\pi L } \epsilon^2 } \sum_{ a = \pm } 
\sum_{ m = 0 }^\infty \biggl\{ \frac{(m+1)(m+2)}{ ( m + 2 )!}   
\left( \frac{g \vpl \epsilon}{
2 L } \right)^2 
\times
\]

\[
\langle a , \{ 0 \} |(\beta P)^2 |  a , \{ 0 \}  \rangle
\prod_{ i = 1}^m   \int_0^\beta \: d \tau_i \: 
\langle a , \{ 0 \} | \tilde{H}_B ( i \tau_j ) 
 |  a , \{ 0 \}  \rangle \biggr\} 
\]

\[
= \tilde{\bf Z}_D [ \pi ]  e^{ - \beta \frac{ g \vpl }{ 4
\pi L } \epsilon^2 } \sum_{ m = 0 }^\infty \biggl\{ 
\frac{(m+1)(m+2)}{ ( m + 2 )!}  ( \beta Y )^{m + 1} [ 1 
+ ( - 1 )^m ] 
\left( \frac{g \vpl }{
2 L } \right)^2 \frac{ \beta \epsilon^2}{ 2 Y } \biggr\} 
\]

\beq 
= \tilde{\bf Z}_D 
[ \pi ] \sum_{ a= \pm } \exp \left[ a \beta \left( Y + \left( 
\frac{g \vpl}{ 2 L } \right)^2 \frac{\epsilon^2}{2 Y } \right) \right]
 \:\:\:\: .
\label{ade61}
\eneq
\noindent

To the leading order in $\epsilon$, the ground state energy derived from 
Eq.(\ref{ade61}) is given by

\beq
E_{\rm GS} [ \epsilon ] = E_{\rm GS} [ \epsilon = 0 ] - Y -
\left( \frac{g \vpl }{2 L } \right)^2 \frac{ \beta \epsilon^2}{ 2 Y }
+ \frac{ g \vpl }{ 4 \pi L } \epsilon^2
\:\:\:\: . 
\label{ade61.1}
\eneq
\noindent

Eq.(\ref{ade61.1}) may be regarded as the lowest order term in the expansion
of the function $ - \sqrt{ Y^2 + \left( \frac{g \vpl \epsilon}{2 L } \right)^2}
$. Although our approach allows to
compute, in principle, also higher-order corrections, for our 
purposes it is enough to notice that, already to the leading order in
$\epsilon$, one gets that the current behaves as a continuous straight 
line, with no sharp, discontinous, jumps, at $\epsilon = 0 $ (i.e., $\varphi
= \pi$).

\beq
I = e^* \frac{ \partial  E_{\rm GS} [ \epsilon ]}{ \partial \epsilon}
\approx  \frac{ g  \vpl }{ 2 \pi L } \epsilon -  \left( \frac{g \vpl }{
2 L } \right)^2 \frac{ \epsilon }{ Y }
\:\:\:\: .
\label{ade63}
\eneq
\noindent

Eq.(\ref{ade63})  is not only consistent with the results of 
previous analyses of rf-SQUIDS  \cite{glazhek,matveev}, but also with a 
strong coupling expansion of the ground state energy of the two-boundary
sine Gordon model derived in Ref.\cite{twoboundary} using the Bethe 
ansatz approach, once  one of the two couplings is set to $\infty$, 
and  a strong coupling expansion in inverse powers of the
other coupling has been performed. 
This makes us confident that the smoothing effect of the
Josephson current due to instantons -exactly computed in 
Ref.\cite{twoboundary} for $g=1$- is effective for any value of the
Luttinger parameter $g$.

\section{The dc-SQUID}

A dc-SQUID is realized as a superconducting loop, interrupted by two
Josephson junctions, of nominal values $E_{J,1}$ and $E_{J,2}$. It
is possible to generalize   to the dc-SQUID the renormalization
group analysis developed  so far for the rf-SQUID, obtaining that, also
in this case, the only alleged fixed points are a weakly coupled
(Neumann), and a strongly coupled (Dirichlet) one.

\begin{figure}
\includegraphics*[width=1.0\linewidth]{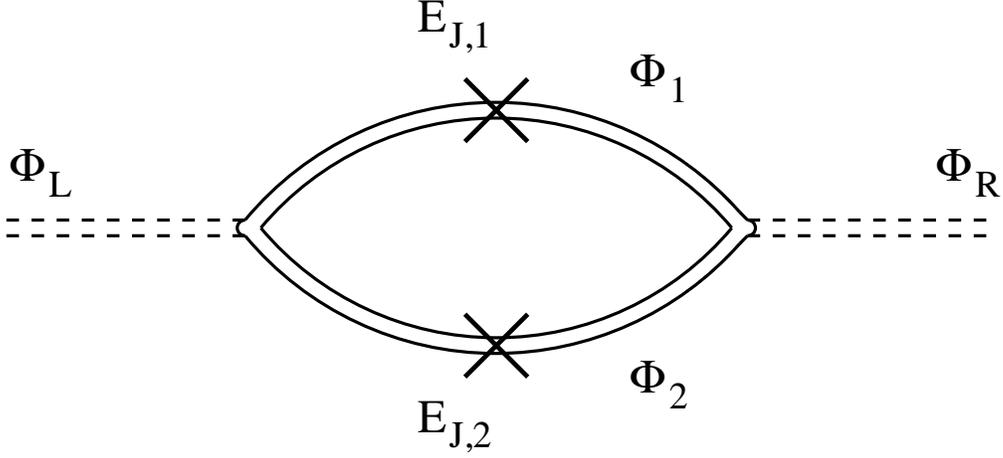}
\caption{Sketch of the dc-SQUID device with the leads.}
\label{figone}
\end{figure}

The device is drawn in Fig.(\ref{figone}): it is made out
of two superconducting wires, connected by two Josephson junctions,
of nominal values $E_{J,1}, E_{J,2}$, with a magnetic flux $\Phi$
piercing the loop. $\Phi_1 ( x )$, $\Phi_2 ( x )$ are the plasmon
fields in the arms connecting the two junctions. 

 When, in the following section, we investigate
the transport properties of a dc-SQUID, the leads to which the device
drawn in  Fig.(\ref{figone}) is connected, are described  by two additional 
plasmon fields  $\Phi_L ( x )$  and $\Phi_R ( x )$. 

The Hamiltonian describing the dynamics of $\Phi_1 (x)$, $\Phi_2(x)$ is 
given by

\beq H_{dc,0} = \frac{g}{2 \pi} \sum_{j=1,2} \int_{-L}^L \: d x \:
\left[ \frac{1}{\vpl} \left( \frac{ \partial \Phi_j }{ \partial t}
\right)^2 + \vpl \left( \frac{ \partial \Phi_j }{ \partial x}
\right)^2 \right] \:\:\:\: ,
\label{edc1}
\eneq
\noindent

while the Josephson energy of the two junctions is described now by the
boundary interaction Hamiltonian $H_{dc,J}$ given by

\beq
H_{dc,J} = - E_{J,1} \cos \left[  \Phi_2 ( L ) - \Phi_1 ( L ) +
\frac{ \varphi}{2}
 \right] - E_{J,2} \cos \left[ \Phi_2 (-  L ) -  \Phi_1 ( -L ) -
\frac{ \varphi}{2} 
 \right]
\:\:\:\: ,
\label{edc2}
\eneq
\noindent
with $\varphi = \Phi / \Phi_0^*$.

Again, it is most convenient to introduce the even- and odd-fields,
$\Phi_{j,e/o} ( x )$, as

\beq
\Phi_{j , e / o } ( x ) =
\frac{1}{ \sqrt{2}} [ \Phi_j ( x ) \pm \Phi_j ( -x ) ]
\:\:\: ; \:\: 0 \leq x \leq L
\:\:\:\: .
\label{edc3}
\eneq
\noindent
By definition, the $\Phi_{j,e} ( x )$'s obey to Neumann boundary conditions at
$x=0$, while the $\Phi_{j,o} ( x )$'s obey to Dirichlet boundary conditions;
namely, one has

\beq
\frac{ \partial \Phi_{1,e} (0 , t)}{ \partial x} = \frac{ \partial \Phi_{2,e}
( 0 , t ) }{\partial x} = 0 \:\:\: ; \:\:
\Phi_{1,o} ( 0 , t ) = \Phi_{2,o} ( 0 ,t ) = 0
\:\:\:\: .
\label{edc4}
\eneq
\noindent

The boundary Hamiltonian $H_{dc , J}$ gets contributions only from 
two linear combinations of the fields $\Phi_e , \Phi_o$. Defining

\beq
X ( x ) = \frac{1}{ \sqrt{2}} [ \Phi_{2,e} ( x ) - \Phi_{1,e} ( x ) ]
\:\:\: ; \:\:
\xi ( x ) = \frac{1}{ \sqrt{2}} [ \Phi_{2,o} ( x ) - \Phi_{1,o} ( x ) ]
\:\:\:\: ,
\label{edc4bis}
\eneq
\noindent

it is easy to realize that the device is  described 
by the reduced Hamiltonian $H_{dc}$  given by

\[
H_{dc} = \frac{g}{ 2 \pi} \int_0^L \: d x \: \left\{
\left[ \frac{1}{\vpl} \left(  \frac{ \partial X}{ \partial t} \right)^2
+ \vpl \left(  \frac{ \partial X }{ \partial x} \right)^2 \right] +
\left[ \frac{1}{\vpl} \left(  \frac{ \partial \xi}{ \partial t} \right)^2
+ \vpl \left(  \frac{ \partial \xi }{ \partial x} \right)^2 \right] \right\}
\]

\beq
- E_{J,1} \cos \left[ X ( L ) + \xi ( L ) + \frac{ \varphi}{2} \right]
- E_{J,2} \cos \left[ X ( L ) - \xi ( L ) - \frac{ \varphi}{2} \right]
\:\:\:\: .
\label{edc5}
\eneq
\noindent

$H_{dc}$ in Eq.(\ref{edc5}) may be regarded as the two-field
generalization of the two boundary sine Gordon Hamiltonian describing the
properties of an rf-SQUID. By construction, at the inner boundary, $x=0$,
$X ( x )$ obeys to Neumann boundary conditions, $\frac{
\partial X ( 0 , t) }{\partial x} = 0$,
while $\xi (x)$ obeys to Dirichlet boundary conditions, $\xi (0 , t) =
0 $. The boundary conditions at the outer boundary, $x=L$, are
obtained by requiring that the energy functional is conserved; namely, by

\[
\frac{ g \vpl}{ \pi} \frac{ \partial X ( L , t)}{ \partial x} + E_{J,1}
\sin \left[ X ( L , t ) + \xi ( L ,t ) + \frac{ \varphi}{2} \right]
+  E_{J,2} \sin \left[ X ( L , t ) - \xi ( L , t) - \frac{ \varphi}{2} \right]
=0 \:\:\:\: ,
\]
\noindent

\beq
\frac{g \vpl}{ \pi} \frac{ \partial \xi ( L , t )}{ \partial x} + E_{J,1}
\sin \left[ X ( L , t ) + \xi ( L , t ) + \frac{ \varphi}{2} \right]
-  E_{J,2} \sin \left[ X ( L , t ) - \xi ( L , t ) - \frac{ \varphi}{2} \right]
=0
\:\:\:\: .
\label{edc6}
\eneq
\noindent

For $E_{J,1} = E_{J,2} = 0$, $X$ and $\xi$  obey to Neumann boundary 
conditions at
$x = L$: $\frac{ \partial X (L , t )}{ \partial x} = \frac{ \partial
\xi (L , t)}{ \partial x} = 0$. For $E_{J,1}, E_{J,2} \to \infty$,
the fields obey to Dirichlet boundary conditions, $X ( L , t ) + \xi (L
, t) - \frac{\varphi}{2} = 0 $ (mod $2 \pi$),  $X ( L  , t ) - \xi
(L , t) + \frac{\varphi}{2} = 0 $ (mod $2 \pi$).

In the following, we shall use $H_{dc}$ to derive the
renormalization of the Josephson energies and the functional form of
the Josephson current in both the weakly coupled and the strongly
coupled regimes accessible to a dc-SQUID.

\subsection{The weakly coupled fixed point}

At the fixed point $E_{J,1} = E_{J,2} = 0 $, $X ( x )$ obeys to Neumann
boundary conditions at both boundaries, while $\xi ( x )$ satisfies
mixed boundary conditions (namely,  Dirichlet boundary conditions at
$x=0$ and Neumann boundary conditions at $x=L$). The mode expansion
of the field $X ( x , t )$ is, then, given by

\beq
X ( x , t ) = 
 q_X + \frac{ \pi \vpl t}{L} \frac{ \tilde{P}_X }{g} +  i 
 \sqrt{\frac{1}{g}}  \sum_{ n \neq 0}
\frac{ \alpha_X ( n ) }{n} \cos  ( k_n x ) \exp (- i k_n  \vpl t )
 \:\:\:\: , \label{eqn.4.abe} \eneq \noindent with $ [
\alpha_X ( n ) , \alpha_X ( n' ) ] = n \delta_{ n + n' , 0 }$; $q_X$
is the constant zero-mode operator and  $k_n = \frac{ \pi}{L} n$.
$\tilde{P}_X$ is, of course,  the "dual" zero-mode operator, whose eigenvalues
are the winding numbers.

The mixed boundary conditions for the field $\xi ( x , t)$ are satisfied by
setting

\beq
\xi ( x , t ) =  -  \sqrt{\frac{1}{g }} \sum_{ n \neq 0 }
\frac{ a_\xi ( n ) }{n } \sin [ K_n x ] e^{ i K_n \vpl t }
\:\:\:\: ,
 \label{eqn.3.abe}
\eneq \noindent with $ [ \alpha_\xi ( n ) , \alpha_\xi ( n' ) ] = n
\delta_{ n + n' , 0 }$, and $K_n = \frac{ \pi}{L} \left( n +
\frac{1}{2} \right)$.

By taking into account Eqs.(\ref{eqn.4.abe},\ref{eqn.3.abe}), the
boundary interaction may be normal ordered as in section 2. In
particular, one obtains

\beq
H_{dc,J} =  - \bar{E}_{J,1} : \cos \left[ X ( L ) + \xi (L)
  + \frac{ \varphi}{2} \right] :
 -\bar{ E}_{J,2} : \cos \left[X ( L ) - \xi (L)
  - \frac{ \varphi}{2} \right] : \:\:\:\: , \label{edc7}
\eneq \noindent with $\bar{E}_{J,j} =  \left( \frac{ 2 \pi a}{L}
\right)^\frac{1}{g} E_{J,j}$, $j=1,2$.

Again, the scaling equations for the pertinent running couplings are
obtained from the requirement that  the partition function $ {\bf Z}$ is
independent of the cutoff. ${\bf Z}$ is given by

\beq {\bf Z}  = \int {\bf D} X {\bf D} \xi \{ e^{ - S_0} {\bf T}_\tau [ e^{
- S_B} ] \} \:\:\:\: ,
\label{eq4.7} \eneq \noindent

with

\[
S_{0}  = \frac{g}{2 \pi} \int_0^\beta \: d \tau \: \int_0^L
\: d x \: \left[ \frac{1}{\vpl} \left(
\frac{ \partial X ( x , i \tau )}{ \partial
\tau} \right)^2 + \vpl \left( \frac{ \partial X ( x , i \tau )}{ \partial x}
\right)^2 \right]
\]

\beq
+ \frac{g}{2 \pi} \int_0^\beta \: d \tau \: \int_0^L \: d x \:
\left[ \frac{1}{\vpl} \left( \frac{ \partial \xi ( x , i \tau )}{
\partial \tau} \right)^2 + \vpl \left( \frac{ \partial \xi ( x , i \tau
)}{ \partial x} \right)^2 \right] \:\:\:\: , \label{eq4.7b} \eneq
\noindent

and

\[
S_B = - \frac{\bar{E}_{J,1}}{2} \int_0^\beta \: d \tau \:
[ e^{ i \varphi} V_{1,1} ( i \tau ) +  e^{ - i \varphi} V_{ -1,-1} ( i \tau ) ]
\]

\beq
 - \frac{\bar{E}_{J,2}}{2} \int_0^\beta \: d \tau \:
 e^{ i \varphi} [ V_{1, - 1} ( i \tau ) +  e^{ - i \varphi} V_{ -1,1} 
( i \tau ) ]
\:\:\:\: ,
\label{eq4.8}
\eneq
\noindent

while the vertex operators $V_{a,b} ( \tau )$ are now given by \beq V_{a,b} (
i \tau ) = : \exp \left\{ i  \left[ b X ( i \tau )
+ a \xi ( i \tau ) \right] - a \frac{ \varphi}{2} \right\} : \;\;\;
; \;\; a,b = \pm 1 \:\:\:\: . \label{eq4.9} \eneq \noindent

To determine the scaling dimensions of the relevant operators, one
needs to compute the correlators of the vertices written in
Eq.(\ref{eq4.9}). They can be computed as discussed  in appendix B;
however, for the purpose of deriving the perturbative RG equations,
one only needs the short-distance limit of the vertex correlators
given by

\beq \langle {\bf T}_\tau [ V_{a,b} ( i \tau ) V_{a',b'} ( i \tau' ) ]
\rangle \approx \left[ \frac{  \pi \vpl}{L} | \tau - \tau' |
\right]^\frac{ ( a a' -1 )}{g} \delta_{b+b' , 0} \:\:\:\:
.\label{eq4.9b} \eneq \noindent Eq.(\ref{eq4.9b}) shows that the
scaling dimension of $V_{a,b}$ is $h_{a, b} = 1 / g$.

Since $ \xi ( \tau )$ does not have a zero mode, the
O.P.E.'s  generate, just as it happens for an rf-SQUID,
  higher periodicity terms, according to
 
\beq V_{a,b} ( i \tau ) V_{a',b'} ( i \tau' ) \approx_{ \tau' \to
\tau^-} \left[ \frac{ \pi \vpl }{L} | \tau - \tau | \right]^{ -
h_{a,b} - h_{a',b'} + h_{a+a' , b+b'}}  V_{a+a',b+b'} ( i \tau' )
 \:\:\:\: .
\label{eq5.3} \eneq \noindent To the second order in the couplings 
$\bar{E}_{J,1} , \bar{E}_{J , 2}$, 
Eq.(\ref{eq5.3}) requires to add to $S_B$ an extra counterterm given
by

\beq
\delta S_B = - \lambda_2 \int_0^\beta \: d \tau \: [  V_{2 , 0} ( i \tau ) +
  V_{- 2 , 0} ( i \tau ) ]
\:\:\:\: .\label{eq5.2} \eneq \noindent

The running coupling constants $G_1,G_2$ are now defined as

\beq G_1 (L) = \left( \frac{L}{ 2 \pi a} \right)^{1 - \frac{1}{g}}
\: \frac{a E_{J,1}}{\vpl} \;\;\; ; \;\; G_2 (L) = \left( \frac{L}{ 2
\pi a} \right)^{1 - \frac{1}{g}} \: \frac{a E_{J, 2}}{\vpl} \:\:\:\:
. \label{eq5.2a} \eneq \noindent

Since the vertices $ V_{ \pm 2 , 0 } ( i \tau ) = e^{\pm i \varphi} :
\exp \left[ \pm  2 i  \xi ( \tau ) \right]:$ are
less relevant than the $V_{\pm 1 , \pm 1}$'s, they do not contribute to the
evaluation of the leading terms of the perturbative 
 $\beta$ functions for  $G_1 , G_2$,

\[
\frac{ d \ln G_1 ( L)}{ d \ln ( L / L_0 )} = \beta_1 ( G_1 , G_2 ) =
\left[ 1 - \frac{1}{g} \right] G_1 (L)
\:\:\:\: ,
\]
\noindent

and

\beq
\frac{ d \ln G_2 ( L)}{ d \ln ( L / L_0 )} = \beta_2 ( G_1 , G_2 ) =
\left[ 1 - \frac{1}{g} \right] G_2 (L)
\:\:\:\: .
\label{eq5.2b}
\eneq
\noindent

As for the rf-SQUID, one finds that both couplings are irrelevant for
$g<1$; namely, the Neumann fixed point is infrared stable, and
the theory is perturbative in the couplings $\bar{E}_{J,1}$,
$\bar{E}_{J,2}$.

When $g>1$, instead, both couplings $G_j ( L )$ increase, when $L$
increases, as $ ( L / L_0 )^{1 - \frac{1}{g}}$. Thus, just as for the
rf-SQUID, the theory becomes nonperturbative as soon as $L \sim
L_*$, with  $L_* \sim 2 \pi a \left(
\frac{\vpl}{a E_{J,{\rm min}}} \right)^\frac{g}{g-1}$, being $
E_{J,{\rm min}}$ the smaller coupling, between $E_{J,1}$ and
$E_{J,2}$.

To evaluate the Josephson current for $g<1$, we compute the partition 
function for finite $\varphi$. The result is

\[
{\bf Z} [ \varphi ] = {\bf Z}_0 \langle {\bf T}_\tau \biggl\{ 
\exp \biggl[ \int_0^\beta \: d \tau \: \biggl[ 
 \bar{E}_{J,1} \cos \left[ X ( i \tau ) + \xi ( i \tau )  
+ \frac{ \varphi}{2} \right]  
\]

\beq
+  \bar{E}_{J,2} \cos \left[  X ( i \tau ) - \xi ( i \tau )
 - \frac{ \varphi}{2} \right] \biggr]  \biggr\} \rangle
\:\:\:\: ,
\label{bc11}
\eneq
\noindent

with

\beq
 {\bf Z}_0 = \prod_{ n = 0}^\infty \biggl\{ 
\left[ \frac{1}{1 - q^{ n + \frac{1}{2} } } \right] 
\left[ \frac{1}{1 - q^{ n + 1 } } \right] \biggr\}
\sum_{ m \in Z } e^{ - \beta \frac{  2 g  \pi \vpl}{L}  m^2 }
\:\:\:\: , 
\label{bc12}
\eneq
\noindent

and $q = e^{ - \beta \frac{ \pi \vpl }{L} }$.
 
Since the field $X ( i \tau )$ has a constant zero-mode term, a perturbative
expansion in the boundary interaction yields nonzero  contributions to the 
partition function  starting only from the second order in $\bar{E}_{ J , 1},
\bar{E}_{J , 2 }$, implying that, at low temperatures,
the leading  $\varphi$-dependence  is  given by

\[
{\bf Z} [ \varphi ] = {\bf Z}_0 \langle \biggl\{ 1 + \frac{1}{2} 
\int_0^\beta \: d \tau_1 \: \int_0^\beta \: d \tau_2 \: 
{\bf T}_\tau [ H_{dc , J} ( \tau_1 ) H_{ dc , J } ( \tau_2 ) ] 
\biggr\} \rangle 
\]

\[
= {\bf Z}_0 \langle \biggl\{ 1 + \beta\bar{E}_{J,1} \bar{E}_{J,2}
\cos ( \varphi ) 
\int_0^\beta \: d \tau \: \frac{ 1}{ \left[ 2 \cosh \left( \frac{ \pi u }{4 
L} \tau \right) \right]^\frac{2}{g} } + \ldots \biggr\}
\]

\beq
 = {\bf Z}_0 \langle \{ 1 + \frac{2 L }{ \pi u } 
\beta \bar{E}_{J,1} \bar{E}_{J,2} B \cos ( \varphi ) \}
\:\:\:\: , 
\label{bc13}
\eneq
\noindent

with 

\beq
B = \frac{ \sqrt{\pi} \Gamma [ \frac{1}{g} ] }{2^\frac{2}{g} \Gamma [ 1 + 
\frac{1}{g} ] } 
\:\:\:\: ; 
\label{bc140}
\eneq
\noindent

as a result, the Josephson current is given by

\beq I [ \Phi ] = - \frac{1}{c \Phi_0^*} \lim_{ \beta \to \infty}
\frac{\partial \ln  {\bf Z} [  \varphi ] }{ \partial \varphi}
 = e^* \frac{2 L }{ \pi u } B
 \bar{E}_{J,1} \bar{E}_{J,2}  \sin ( \varphi )
\:\:\:\: . 
\label{bc14}
\eneq
\noindent

The functional form of $ I [ \Phi ]$ is similar to
Eq.(\ref{eqaddi1.10}), but it is now proportional to  
$ \bar{E}_{J,1} \bar{E}_{J,2}$. For
$g<1$, Eq.(\ref{bc14}) is valid for any value of $L$, since the
quantum corrections are perturbative in $ \bar{E}_{J,1} \bar{E}_{J,2}$. 
When $g>1$, instead,
higher harmonics operators become more and more relevant as $L$
increases, inducing a crossover to a sawtooth-like functional form
of the Josephson current, as  $L$ approaches  $L_*$.

\subsection{The strongly coupled fixed point}

In this subsection the strong coupling ($E_{J,j} / ( 2 \pi \vpl / L
) \ll 1$) regime of a dc-SQUID is analyzed and the explicit
functional forms of the partition function and of the Josephson
current are derived. To compute the  partition function with
Dirichlet-like boundary conditions at the outer boundary, $ X ( L ,
t ) + \xi ( L , t ) + \frac{ \varphi}{2} = 0$ (${\rm mod } \: 2
\pi)$), and $X ( L , t ) - \xi ( L , t ) - \frac{ \varphi}{2} = 0$
(${\rm mod } \: 2 \pi)$), one notices that

\beq
 X ( L , t ) = \pi (n_1 + n_2 ) \;\;\; ; \;\; \xi ( L , t ) -
\frac{\varphi}{2} = \pi (n_1 - n_2 ) \:\:\:\: . \label{eqn.2.a}
\eneq \noindent with $n_1 , n_2$ relative integers.

Eqs.(\ref{eqn.2.a}), together with the boundary conditions at the
inner boundary, lead to

\beq
\xi ( x , t ) =  - \frac{2 \pi}{L} P_\xi x - 
\sqrt{\frac{1}{g}} \sum_{ n \neq 0 }
\sin \left[ \frac{ \pi n }{L} x \right] \frac{ \alpha_\xi ( n ) }{ n } 
e^{ - i  \frac{ \pi n }{L} ut}
\:\:\:\: ,
 \label{eqn.3.a}
\eneq \noindent with $ [ a_\xi ( n ) , a_\xi ( n' ) ] = n \delta_{n
+ n' , 0 } $, $k_n =  \pi n / L$; the 0-mode operator of the field $\xi$,
$P_\xi$, has the eigenvalues

\beq 
P_\xi =  \left[ \frac{ \varphi}{4 \pi } +
\frac{ m }{2}  \right]  \:\:\: ; \:\: ( m = n_1 - n_2 ) \:\:\:\: .
\label{eqn.3.aultra} 
\eneq 
\noindent

Similarly, one  obtains

\beq
X ( x , t ) = q_X + i \sqrt{\frac{1}{g}} \sum_{ n \in Z} \cos \left[ 
\frac{\pi}{L}
\left( n + \frac{1}{2} \right) x \right] \frac{ \alpha_X ( n ) }{ 
n + \frac{1}{2} } e^{ - i \frac{\pi}{L} \left( n + \frac{1}{2} \right) u t }
\:\:\:\: , \label{eqn.4.a} \eneq \noindent with $ [ \alpha_X ( n ) ,
\alpha_X ( n' ) ] = n \delta_{ n + n' , 0 }$, $K_n = \frac{ \pi}{L}
\left( n + \frac{1}{2} \right)$, $q_X = \sqrt{ \frac{g}{2}} \pi (
n_1 + n_2 )$.

As for the rf-SQUID, the strongly
coupled fixed point is described in terms of the (now two!) dual fields 
$\Theta_X ( x ,t ) $ and $ \Theta_\xi ( x , t )$, whose mode expansion is 
given by

\beq
 \Theta_X ( x , t ) = - 2 \pi \tilde{P}_X - 2 \sqrt{ g}
\sum_{ n \in Z} \sin \left[ \frac{\pi}{L} \left( n + \frac{1}{2} \right) x
\right]  \frac{ \alpha_X ( n ) }{ n + \frac{1}{2} }   
e^{ - i \frac{\pi}{L} \left( n + \frac{1}{2} \right) u t } 
\:\:\:\: , 
\label{coc4}
\eneq
\noindent

where $P_X$ is the canonical conjugate operator of $q_X$, and

\beq
\Theta_\xi ( x , t ) = \theta_0 + \frac{ 2 \pi}{L} g P_\xi
u t + i 2 \sqrt{ g } \sum_{ n \neq 0  } \cos \left[ \frac{\pi n}{L} x \right]
 \frac{ \alpha_\xi ( n ) }{n} e^{ - i \frac{ n \pi}{L} u t } 
\:\:\:\: . 
\label{coc5}
\eneq
\noindent

Following the same steps used in  subsection  (2.2), one has that 

\beq
H_D [ \Theta_X , \Theta_\xi ] = \frac{ 2 \pi \vpl  g }{L} ( P_\xi )^2 + 
\sum_{ n \neq 0 } [ \alpha_X ( -n   ) \alpha_X ( n - 1 ) + 
\alpha_\xi ( - n ) \alpha_\xi ( n ) ] 
\:\:\:\: ,
\label{coc6}
\eneq
\noindent

from which it is straightforward to compute the partition 
function at the Dirichlet fixed point:

\beq
{\bf Z}_D = {\rm Tr} [ e^{ - \beta H_D [ \Theta_X , \Theta_\xi ] } ] = 
\prod_{ n = 0}^\infty \left[ \left( \frac{1}{1 - q^{ n + \frac{1}{2} } }
\right) \left( \frac{1}{1 - q^{ n + 1 } } \right) \right]
\sum_{ k \in Z } \exp \left[ - \beta \frac{ \pi \vpl g}{ 2  L } \left( k
- \frac{ \varphi}{2 \pi} \right)^2 \right]
\:\:\:\: .
\label{coc7}
\eneq
\noindent

Eq.(\ref{coc7}) shows that, for $\varphi \neq 2 k \pi + \pi$, with
$k$ integer, the ground state is characterized by a nonzero
value of $P_\xi$, proportional to $[\varphi]$. Tunneling events from this
ground state to the nearest (in energy) ground states 
are suppressed by the  exponential factor $
\exp [ - \beta  \pi \vpl g / 2 L ]$.

Taking the logarithmic derivative of Eq.(\ref{coc7}) one easily 
derives the Josephson current, which is given by

\beq
I [ \varphi ] = - e^* \lim_{ \beta \to \infty } \frac{1}{\beta} \frac{ \partial
\ln {\bf Z}_D [ \varphi ] }{ \partial \varphi } 
= \frac{ e^* g \vpl }{ 4 \pi L } \{ \varphi - [ \varphi ] \}
\:\:\:\: . 
\label{coc8}
\eneq
\noindent

From Eq.(\ref{coc7}), one sees that the degeneracy among the minima of the
Josephson energy is removed by the magnetic energy $E^{(0)}_m =
\frac{ 2 g \pi \vpl }{L} \left( - \frac{\varphi}{4 \pi} + \frac{m}{2}
\right)^2$. Since  $E^{(0)}_m $ = $E^{(0)}_{m
+ 1} $, for $\varphi = \pi + 2 \pi m$ also a dc-SQUID
may operate as a qubit between these two levels. When the couplings
are large, but finite, phase slips in the plasmon field (instantons)
will induce  a  renormalization of the physical parameters, as we shall
show in the following subsection.

\subsection{Instantons and boundary perturbations at the Dirichlet fixed point}

In order to construct the leading boundary perturbations at the strongly 
coupled fixed point, one may use again the DEBC-method 
\cite{wires}, applied to the primary bulk operators involving the dual
fields $\Theta_X , \Theta_\xi $. A generic, imaginary time, bulk operator 
at the strongly coupled fixed point may be written as
$V_{\alpha , \beta } ( i \tau ) = : \exp \{ i [ \alpha \Theta_X ( i \tau ) + 
\beta \Theta_\xi ( i \tau ) ] \}$, where the imaginary time boundary fields 
are given by

\beq
 \Theta_X ( i \tau   ) = \Theta_X ( L , i \tau  ) = - 
2 \pi \tilde{P}_X - 2 \sqrt{ g} \sum_{ n \in Z} ( -1 )^n 
\frac{ \alpha_X ( n ) }{ n + \frac{1}{2} }   
e^{ -  \frac{\pi}{L} \left( n + \frac{1}{2} \right) u \tau } 
\:\:\:\: , 
\label{coc4i}
\eneq
\noindent

 and by

\beq
\Theta_\xi ( i \tau  ) = \Theta_\xi ( L , i \tau ) = 
\theta_0 + \frac{ 2 \pi}{L} 2 g P_\xi i 
u \tau  + i 2 \sqrt{ g } \sum_{ n \neq 0  } ( -1 )^n 
\frac{ \alpha_\xi ( n ) }{n}
e^{ -  \frac{ n \pi}{L} u \tau } 
\:\:\:\: . 
\label{coc5i}
\eneq
\noindent

From the commutation rules between the zero modes of the two
fields, one obtains

\beq
[ q_X , V_{\alpha , \beta } ( i \tau )  ] = 2 \pi \alpha V_{\alpha , \beta } 
( i \tau )  \:\:\: ; \:\:
[ P_\xi , V_{\alpha , \beta } ( i \tau )  ] = 2 \pi \beta
V_{\alpha , \beta } ( i \tau ) 
\:\:\:\: . 
\label{coc5.1}
\eneq
\noindent

Thus, $V_{\alpha , \beta } ( i \tau ) $ changes the eigenvalue of $q_X$ by
$2 \pi \alpha$ and the eigenvalue of $P_\xi$ by $2 \pi \beta$. 
Using Eqs.(\ref{coc4i},\ref{coc5i},\ref{coc5.1}), one may write the 
leading boundary operators as

\beq
\tilde{V}_{ a , b } ( i \tau ) = : \exp \left\{ \frac{i}{2} [ 
a \Theta_X ( i \tau )  + b \Theta_\xi ( i \tau )] 
\right \} \;\;\; ; \;\; a , b  = \pm 1
\:\:\:\: , 
\label{coc9}
\eneq
\noindent

\beq
\tilde{V}_{ X , \pm } ( i \tau ) = 
 : \exp \{ \pm i \Theta_X ( i \tau ) \} : \;\;\;  ;
\;\; \tilde{V}_{ \xi , \pm } ( i \tau ) =  
: \exp \{ \pm i \Theta_\xi ( i \tau ) \}
\:\:\:\: .
\label{cocic1}
\eneq
\noindent

$\tilde{V}_{ 1 , 1 } ( i \tau )$ and  
$\tilde{V}_{ - 1 , - 1 } ( i \tau )$ ($\tilde{V}_{ 1 , - 1 } ( i \tau )$ and  
$\tilde{V}_{ - 1 ,  1 } ( i \tau )$) change by $\pm 2 \pi$ the phase at
junction 1 (2), while they leave the phase at junction 2 (1) unchanged and
thus they describe phase slip operators at junction 1 (2). Similarly, 
 $\tilde{V}_{ X , \pm } ( i \tau )$ ($\tilde{V}_{ \xi , \pm } ( i \tau )$)
change the eigenvalue of the overall (relative) phase $ X ( L )$  
($\xi ( L )$) by $\pm 2 \pi$ leaving the eigenvalue of $\xi ( L )$
($ X ( L )$) unchanged: thus, $\tilde{V}_{ X , \pm } ( i \tau )$  do not
affect the Josephson current across the ring, while 
 $\tilde{V}_{ \xi , \pm } ( i \tau )$  provide corrections to the Josephson 
current when  $\varphi = 2 k \pi + \pi$. In the following we shall refer to
$\tilde{V}_{ X , \pm } ( i \tau )$ ($\tilde{V}_{ \xi , \pm } ( i \tau )$)
as $q$- ($\pi$- ) Instanton operators. 

The leading boundary perturbation at the strongly coupled fixed point is
given by an arbitrary linear combinations of the operators listed above, as

\[
\tilde{H}_{dc , J } = Y_{1} [ \tilde{V}_{1,1} ( i \tau ) + 
 \tilde{V}_{ -1 , - 1} ( i \tau ) ] 
+ Y_{2} [ \tilde{V}_{1, - 1} ( i \tau ) + 
 \tilde{V}_{ -1 ,  1} ( i \tau ) ] 
\]

\beq
+  Y_{X} [ \tilde{V}_{ X , + } ( i \tau ) :  + 
\tilde{V}_{ X , - } ( i \tau ) ] + 
 Y_{\xi} [\tilde{V}_{ \xi , + } ( i \tau ) :  + 
\tilde{V}_{ \xi , - } ( i \tau ) ] 
\:\:\:\: . 
\label{coco2}
\eneq
\noindent

From the mode expansion of the dual fields, it is straightforwad to compute
the (low-temperature) correlators among the boundary vertex operators
appearing  in  Eq.(\ref{coco2}). For $ | \varphi | < \pi$, for instance, 
one has that 

\beq
\langle {\bf T}_\tau [ \tilde{V}_{ 1 , 1 } ( i \tau ) \tilde{V}_{ - 1 , - 1 } 
( i \tau'  ) ] \rangle =
\langle {\bf T}_\tau [ \tilde{V}_{ - 1 , 1 } ( i \tau ) \tilde{V}_{  1 , - 1 } 
( i \tau'  ) ] \rangle = 
\frac{e^{ \frac{ \pm g \vpl \varphi}{4 L } (\tau  - \tau' ) }}{ 
\left|  \sinh \left[ \frac{ \pi \vpl}{4 L } ( \tau - \tau' ) 
\right] \right|^{2g} }
\:\:\:\: , 
\label{corcoc1}
\eneq
\noindent

\beq
\langle {\bf T}_\tau [ \tilde{V}_{ + , X} ( i \tau ) \tilde{V}_{ - , X } 
( i \tau'  ) ] \rangle = 
\frac{\left|  \cosh \left[ \frac{ \pi \vpl}{4 L } ( \tau - \tau' ) 
\right] \right|^{4 g}}{ 
\left|  \sinh \left[ \frac{ \pi \vpl}{4 L } ( \tau - \tau' ) 
\right] \right|^{4 g} }
\:\:\:\: , 
\label{corcoc2}
\eneq
\noindent

and 

\beq
\langle {\bf T}_\tau [ \tilde{V}_{ + , \xi } ( i \tau ) \tilde{V}_{ - , \xi } 
( i \tau'  ) ] \rangle = 
\frac{e^{ \pm \frac{ g \vpl \varphi}{ L } (\tau  - \tau' ) }}{ 
\left| 2 \sinh \left[ \frac{ \pi \vpl}{2 L } ( \tau - \tau' ) 
\right] \right|^{4 g} }
\:\:\:\: . 
\label{corcoc3}
\eneq
\noindent

Eqs.(\ref{corcoc1},\ref{corcoc2},\ref{corcoc3}) imply  that 
 the fugacity of the q- and $\pi$-instantons and of the phase slips
at the junctions scale with the SQUID size as

\beq
Y_{ X ( \xi ) } ( L ) = Y_{ X ( \xi ) }  ( \vpl T_x )
\left( \frac{L}{\vpl T_x} \right)^{1 - 2 g} \:\:\: ; \:\:
Y_{1 ( 2 ) } ( L ) = Y_{ 1 ( 2 ) }  ( \vpl T_x )
\left( \frac{L}{\vpl T_x} \right)^{1 -  g} \:\:\:\: . 
\label{corcoc4}
\eneq
\noindent

The "instanton size" $T_x$ in Eq.(\ref{corcoc4}) 
is estimated to be of the order of 
 $\sqrt{ g L / [ \pi \vpl (  E_{J,1} + E_{J,2} ) ] }$ \cite{glazhek}.

Eq.(\ref{corcoc4}) shows that, for $g<1$, phase slips at the two junctions
 provide the most relevant perturbation to the Dirichlet fixed point, 
rendering the strongly coupled fixed point  not infrared stable. 
For finite $\beta , 
L$ and for $\varphi \neq 2 \pi k + \pi$, however, the phase slips  are always 
suppressed by the  magnetic energy associated to the eigenvalue of $P_\xi$.
At variance, when $g>1$, $\pi$-instantons are an irrelevant perturbation 
which, 
for $L < L_{*,1} , L_{*,2}$ (i.e., of the healing lengths associated to the 
two junctions), only smooths down the edges
of the sawtooth shape of the Josephson current: for $L
\to \infty$, the ground state exhibits a discrete 
Z-symmetry.

\section{Transport in SQUID devices}

To probe the different regimes attainable by SQUID devices, it is
most useful to look at their dc-conductance. To get  conduction of
current across a SQUID device,  one  should be able to
connect it to two leads, enabling to apply a
biasing voltage $V$. This can be achieved easily for
a dc-SQUID while, for the purposes of this section, it is most
convenient to regard the rf-SQUID as an inhomogeneous chain of Josephson 
junctions
 connected to two bulk superconducting leads at fixed phase difference 
$\varphi$ and at finite biasing voltage $V$; the chain is made by
junctions of equal strength with a weak link of nominal strength 
$E_J$ located at its center.  For $V=0$, this 
inhomogeneous chain mimicks the response of an rf-SQUID to an external 
magnetic flux  $\varphi$ \cite{GiuSo}. 
In  this section we  study the trasport properties of both devices,
evidencing the different current (normal and Josephson) response at  weak
and at strong coupling. Our boundary field theory approach well reproduces
the known results of Refs.\cite{glazhek,zwergerkane} and, for a dc-SQUID, 
evidences interesting interference effects between the current flowing 
through the two junctions.

\subsection{The Josephson junction chain at the weakly coupled fixed point}

The relevant dynamics of a Josephson junction chain with a weak link connected
to two bulk superconductors at fixed phase difference $\varphi$ has been 
altready analyzed in Refs.\cite{glark,GiuSo}. The properties of this 
device may be  described  by the bosonic field $\Phi ( x ) = 
\frac{1}{ \sqrt{2}} [ \Phi_R ( x ) - \Phi_L ( - x ) ]$, where $\Phi_R ( x )$ 
and $\Phi_L ( x )$ are the fields describing the phase of each junction in 
the right and the left half-chain, respectively. The current operator
across the junction is then  given by $j ( t ) = -  \frac{g e^*  \vpl }{  
\sqrt{2} \pi } \frac{ \partial  \Phi ( L , t )}{  \partial x}$. Energy 
conservation provides the dynamical boundary conditions for $\Phi ( x , t )$ 
at the outer boundary, which is given by

\beq
\frac{ \vpl g }{ \pi} \frac{ \partial \Phi ( L , t )}{ \partial x} 
+ \sqrt{2} \bar{E}_J \sin [ \sqrt{2} \Phi  ( L , t )  + \varphi ] =  0 
\:\:\:\: ;
\label{ade9}
\eneq
\noindent
using Eq.(\ref{ade9}), the current operator may be written as 

\beq
j ( t )=  e^* \bar{E}_J \sin [ \sqrt{2} \Phi  ( t )  + \varphi ]
\:\:\: ; \;\; ( \Phi ( t ) = \Phi (L , t )) \;\;\;\: . 
\label{ade10}
\eneq
\noindent

At the weakly coupled fixed point ($E_J = 0$), $\Phi ( x , t )$ satisfies the 
Neumann boundary conditions at the outer boundary; namely, 
$\frac{ \partial \Phi ( L  , t ) }{ \partial x} = 0$.

In order to apply a voltage $V$ to the device, one may bias  at voltage $V/2$ 
and $-V/2$ each one of the two leads.  In the small 
$\bar{E}_J$-limit, it is safe to assume that each chain is at equilibrium with
its own lead, and that the two  half chains are disconnected. 
Accordingly, the voltage bias $V$ is introduced by adding
to the total Hamiltonian  the term

\beq
H_V = - \frac{ e^* V g}{  \pi \vpl} \: \int_0^L \: d x \: 
\frac{ \partial \Phi }{ \partial t} 
\:\:\:\: , 
\label{ade20}
\eneq
\noindent

which can be accounted for by a mere shifting of the field $\Phi$, as

\beq
\Phi ( x , t ) \to \Phi ( x , t ) - e^* V t 
\:\:\:\: . 
\label{ade21}
\eneq
\noindent

As a result of  Eq.(\ref{ade21}), the Josephson current
acquires an explicit dependence on time and, to the first order in
$\bar{E}_J$, reads

\beq
I_J [ \varphi , t ] = e^* \bar{E}_J \sin [ \varphi + e^* V t ]
\:\:\:\: . 
\label{ade2112}
\eneq
\noindent

Thus, in addition to the Josephson current, there is 
a normal time-independent dc-current which, as it will be
shown in the following, becomes the leading contribution for a pertinent
choice of the infrared cutoff leading to the celebrated Kane-Fisher
formula \cite{zwergerkane}. 

Since the normal current is independent of time, it 
may be computed using the imaginary time formalism.
From Eq.(\ref{ade21}), one obtains

\beq
j ( i \tau ) =  e^* \bar{E}_J   : \sin [ \sqrt{2} \Phi  ( i \tau )  + 
\varphi + e^* V i \tau   ]:
\:\:\:\: ,
\label{ade22}
\eneq
\noindent
with

\beq
\Phi ( i \tau ) = - \sqrt{\frac{2}{g}}  \sum_n (-1)^n
\frac{ \alpha ( n ) }{ n + \frac{1}{2} } \: e^{ - 
\frac{ \pi}{L} \left( n + \frac{1}{2} \right) \vpl \tau }
\:\:\:\: . 
\label{ade7}
\eneq
\noindent

From Eqs.(\ref{ade22},\ref{ade7}), it follows that  the leading 
 contribution to the normal current appears only at the second order 
in $\bar{E}_J$, since 

\[
I \approx    \frac{e^*}{4 i }  ( \bar{E}_J )^2 \int_0^\beta \: d \tau' \: 
\langle {\bf T}_\tau [  : \exp [ \sqrt{2} \Phi  ( i \tau )  + \varphi +
i  e^* V \tau  ]: \:  : \exp [ - \sqrt{2} \Phi ( i \tau' )  - \varphi -
i e^* V \tau' ] : ]
\rangle
\]

\[
-  \frac{e^*}{4 i }  ( \bar{E}_J )^2 \int_0^\beta \: d \tau' \: 
\langle {\bf T}_\tau [  : \exp [ - \sqrt{2} \Phi  ( i \tau )  - \varphi -
i  e^* V \tau  ]: \:  : \exp [  \sqrt{2} \Phi ( i \tau' )  + \varphi +
i e^* V \tau' ] : ]
\rangle
\]

\beq
=   e^* ( \bar{E}_J )^2 \: \frac{ 2 L}{\pi \vpl } 
\: \int_0^\infty \: d x \: \left[ \frac{ \sinh \left[ 
a x \right]  }{ 2^\frac{2}{g}  
\sinh^\frac{2}{g} ( x ) } \right] =  
\frac{ 2 e^* \pi ( \bar{E}_J )^2  L }{ \pi \vpl } {\bf I} [ a ]
\:\:\:\: , 
\label{ade24}
\eneq
\noindent
with

\beq
a =  \frac{ 2 L e^* V }{\pi \vpl }
\:\:\:\: . 
\eneq
\noindent

and  \cite{abram}

\beq
{\bf I} [ a ] = \int_0^\infty \: d x \: \frac{ \sinh [  a x ] }{[ 2
\sinh ( x ) ]^\frac{2}{g} } = \frac{\Gamma [ 1 - \frac{2}{g} ] }{4  } 
\biggl\{ \frac{ 
\Gamma [ -  \frac{a}{2} + \frac{1}{g} ] }{ \Gamma [ 1 - 
 \frac{a}{2} - \frac{1}{g} ] } - \frac{ 
\Gamma [   \frac{a}{2} + \frac{1}{g} ] }{ \Gamma [ 1 + 
 \frac{a}{2} - \frac{1}{g} ] } \biggr\}
\:\:\:\: . 
\label{ade25}
\eneq
\noindent

From Eq.(\ref{ade24}), one sees that  the dc-current depends crucially
on  the ratio $a$ between the energy window 
defined by the applied voltage, $e^* V$, and the level spacing 
characteristic of the device,  $\pi \vpl / (2 L) $. 
In fact, if $ e^* V < \pi \vpl / L$, one may
 expand Eq.(\ref{ade24}) to first order in $a$,  getting

\beq
I =   \frac{  e^*  ( \bar{E}_J )^2  L 
\Gamma [ 1 - \frac{2}{g} ] }{ 2 \vpl } 
B \: V 
\:\:\:\: , 
\label{adelante2}
\eneq
\noindent

with 

\beq
B = \frac{\partial }{ \partial z } \biggl\{ \frac{ 
\Gamma [ - \frac{z}{2} + \frac{1}{g} ] }{ \Gamma [ 1 - 
\frac{z}{2} - \frac{1}{g} ] } - \frac{ 
\Gamma [  \frac{z}{2} + \frac{1}{g} ] }{ \Gamma [ 1 + 
\frac{z}{2} - \frac{1}{g} ] } \biggr\} \biggr|_{ z = 0 }
\:\:\:\: ,
\label{adelante25}
\eneq
\noindent

which yields  the (normal) 
dc-linear response, to an applied voltage
$V$, of a junction with renormalized strength $\bar{E}_J = \left( \frac{2 \pi
a}{L} \right)^\frac{1}{g}$; this is expected, since, in this limit,
 $\pi \vpl / L $  provides the infrared cutoff.
If, instead, $ \vpl / L \ll e^* V$,  one
may compute the current in a large-$a$ expansion, using the  Stirling
formula, $\Gamma [ z ] \approx 
\sqrt{2 \pi} e^{ - z } z^{ z - \frac{1}{2}}$, which yields

\beq
I =  \frac{ e^*  ( \bar{E}_J )^2 L }{ \pi \vpl } \sin \left[ \frac{  \pi}{g} 
\right] \Gamma \left[ 1 - \frac{ 2}{g} \right] \left[ \frac{ 2 L e^* V}{ 
\pi \vpl } \right]^{\frac{2}{g} - 1 } 
\:\:\:\: .
\label{ade26}
\eneq
\noindent

Eq.(\ref{ade26}) is the celebrated Kane-Fisher formula for the current
across a constriction in a spinless Luttinger liquid. Our result is, after
all, not surprising, since  the wires connecting
the junctions have been regarded  as  one-dimensional spinless Luttinger 
liquids, and
the junctions as boundary interactions. It should be noticed, however,  that
the normal current is proportional to $ ( e^* V )^{ \frac{2}{g} - 1}$ since
$e^* V$ is the pertinent infrared cutoff when an external bias voltage is 
applied to the device. At variance, in the computation of the Josephson 
current, the pertinent cutoff is provided by the plasmon's level spacing
 $ s = 2 \pi  \vpl  /  L $. Thus, in the limit where $s \ll e^* V$, our
analysis shows that the leading contribution to the total current 
across the junction is  given by the normal current.

\subsection{The Josephson junction chain at the strongly coupled fixed point}

In this section, the current across the Josephson junction chain is
computed at zero phase difference $\varphi$ as
 $\bar{E}_J / ( \pi u / L ) \to
\infty$. Here,  it is most convenient to resort to the dual 
formulation, based on the dual field $\Theta ( x , t )$. The 
total Hamiltonian (including the leading boundary perturbation) is given by

\beq
H [ \Theta ] = \frac{1}{ 4 \pi (2 g)} \: \int_0^L \: d x \: 
\left[ \frac{1}{u} \left( \frac{ \partial \Theta }{ \partial t} \right)^2 + 
u \left( \frac{ \partial \Theta }{ \partial x} \right)^2 \right]
- Y [ : e^{ \frac {i}{\sqrt{2}} \Theta ( L ) }: + 
 : e^{ \frac {i}{\sqrt{2}} \Theta ( L ) }:  ]
\:\:\:\: , 
\label{transps1}
\eneq
\noindent

where the dual field $\Theta ( x , t )$ obeys to Neumann boundary conditions at
$x = L$. Thus, the mode expansion of the field $\Theta ( i \tau ) = 
\Theta ( L , i \tau )$  is given by

\beq
\Theta ( i \tau ) = \Theta ( L , i \tau ) = 
\theta_0 + i \frac{ 2 \pi }{L} 2 g P u \tau 
+ i 2 \sqrt{g } \sum_{ n \neq 0 } (-1)^n  \frac{ \alpha ( n ) }{ n } 
e^{ -  \frac{ \pi n }{L} u \tau } 
\:\:\:\: .
\label{ade36bis}
\eneq
\noindent

Since at the strongly coupled  fixed point, the two half  chains cannot be
regarded as isolated from each other,  one must apply a voltage $V$ by 
biasing the right-handed chiral mode of $\Phi ( x , t )$, with respect to the 
left-handed ones. This  corresponds to adding  to $H [ \Theta ]$ 
a voltage-dependent term given by

\beq
\tilde{H}_V =  - \frac{e^* V  }{  2 \pi  } \int_0^L \: dx \: 
 \frac{ \partial \Phi }{ \partial x} =
 - \frac{e^* V  }{  2 \pi \vpl } \int_0^L \: dx \: 
 \frac{ \partial \Theta }{ \partial t}
\:\:\:\: . 
\label{ade45}
\eneq
\noindent

The imaginary time current operator is 

\beq
j ( i \tau ) =  \frac{ \delta  S_E [ \Theta ; V , a ] }{ \delta a ( \tau )}
\biggr|_{ a = 0 }
\:\:\:\: ,
\label{ade45c}
\eneq
\noindent
where $S_E [ \Theta ; V , a ]$ is the Euclidean action with a  source term
for the current,  given by

\[
 S_E [ \Theta ; V , a ] = \frac{1}{ 4 \pi (2 g) } \: \int_0^\beta \: d \tau \:
\int_0^L \: d x \: \left[ \frac{1}{\vpl} \left( \frac{ \partial \Theta ( x , 
i \tau )}{ \partial \tau } \right)^2 + \vpl  
\left( \frac{ \partial \Theta ( x , 
i \tau )}{ \partial x } \right)^2 \right] 
\]

\[
- 2 Y \: \int_0^\beta
\: d \tau \: : \cos ( \frac{ \Theta ( i \tau ) }{ \sqrt{2}} ):
\]

\beq
 - i \frac{e^* V  }{  2 \pi \vpl } 
\: \int_0^\beta \: d \tau \: \int_0^L \: dx \: 
 \frac{ \partial \Theta ( x , 
i \tau )}{ \partial \tau } - \frac{ i e^* }{ 2 \pi L }
\int_0^\beta \: d \tau \: \int_0^L \: dx  \: a ( \tau ) \frac{ \partial 
\Theta ( i \tau )}{ \partial \tau }
\:\:\:\: ,
\label{ade45b}
\eneq
\noindent

By shifting the field $\Theta$ according to

\beq
\frac{ \Theta ( x , i \tau ) }{ \sqrt{2}}  \to \frac{ \Theta ( x , i \tau ) }{ 
\sqrt{2}} - 2 i g e^* V \tau  - i
\frac{  e^* g \vpl}{L} A (  \tau )
\:\:\:\: , 
\label{ade45d}
\eneq
\noindent

with $A' ( \tau ) = a ( \tau )$, one finds that the
current is given by

\beq
\tilde{j} ( \tau ) =   \frac{ \delta  \tilde{S}_E [ \Theta ; V , a ] }{ 
\delta a ( \tau )} \biggr|_{ a = 0 } = \frac{ g ( e^* )^2 }{ \pi} V 
+  4 Y \frac{  e^* g \vpl}{L}  \sin ( \frac{ \Theta ( i \tau )}{ \sqrt{2}}
  - i  e^* V \tau ) : 
 \:\:\:\: . 
\label{ade46}
\eneq
\noindent

From Eq.(\ref{ade46}) one sees that, apart from the constant 
term $\frac{ g ( e^* )^2 }{ \pi} V $, the current may be computed as 
in the previous section,  provided that one substitutes $4/g$ with
$g$ and  $\bar{E}_J$ with $Y$. Thus, to the second order in $Y$ one
obtains

\beq
I = \frac{ g ( e^* )^2 }{ \pi} V + 
 \frac{2 e^*  Y^2 L }{ \pi  } \sin \left[  g \pi
\right] \Gamma \left[ 1 - 2 g \right] \left[ \frac{ 2 L e^* V}{ 
\pi \vpl } \right]^{ 2 g - 1 } 
\:\:\:\: .
\label{ade47}
\eneq
\noindent

As it happens also at the weakly coupled fixed point, the celebrated 
power-law dependence of $I$ on
the applied voltage $V$ emerges when $2 L e^* V / ( \pi \vpl ) \gg 1$.

\subsection{The dc-SQUID at the weakly coupled fixed point}

For a dc-SQUID, it is most convenient to 
 represent the leads as two quantum wires
with the same Luttinger parameters $g$ and $\vpl$, with plasmon fields 
respectively given by $\Phi_L$ and $\Phi_R$. The two leads are 
assumed to be connected to the dc-SQUID with smooth junctions, that is, 
three-wire junctions with no backscattering in any arm.
Thus, one should  describe the SQUID's arm with
four fields, rather than with two; namely $\Phi_{u,L} , \Phi_{u, R}$ for the
left-upper and for the right-upper arm of the ring, 
 $\Phi_{d,L} , \Phi_{d, R}$ for left-lower and for the right-lower arm of the 
ring, respectively. Since the connections between leads and SQUID's arm are
assumed to be ideal, they are described by the strongly coupled 
fixed point of the three quantum wire "Y"-junction studied in Ref.\cite{wires},
 with no concatenated flux. Thus, by centering the Y-junction
 at $x=0$, the boundary conditions are given by

\beq
\frac{ \partial }{ \partial x} [ \Phi_{ a} ( x , t ) + \Phi_{u , a } ( x , t ) 
+ \Phi_{ d , a } ( x , t ) ] \biggr|_{ x = 0 } = 0
\:\:\:\: , 
\label{scs1}
\eneq
\noindent

which embodies the current conservation at the connection, and by

\beq
-2 \Phi_a ( 0 , t ) + \Phi_{u , L } ( 0 , t ) 
+ \Phi_{ d , a } ( 0 , t ) = \Phi_{u , a } ( 0 , t ) -  
\Phi_{ d , a } ( 0 , t )  = 0 
\:\:\:\: , 
\label{scs2}
\eneq
\noindent
with $a = L , R$.

The relevant fields for the Josephson junction dynamics are given by the 
"odd-parity" combinations 
\beq
\Phi_{ u / d }  ( x , t ) = \frac{1}{ \sqrt{2}} [ \Phi_{ u/d , R } ( x , t ) 
- \Phi_{ u/d , L } ( x , t ) ] \;\;\; ; \;\;
\Phi ( x , t ) = \frac{1}{ \sqrt{2}} [ \Phi_{ R } ( x , t ) 
- \Phi_{ L } ( x , t ) ] 
\:\:\:\: , 
\label{scs4}
\eneq
\noindent

which are  related to the fields 
$\xi ( x , t ) $ , $X ( x , t )$ introduced in section (3) by

\beq
\Phi_{u} ( x , t ) = \frac{1}{2} [ X ( x , t ) + \xi ( x , t ) ]
\;\;\; ; \;\;
\Phi_{d} ( x , t ) = \frac{1}{2} [ X ( - x , t ) - \xi ( - x , t ) ]
\:\:\:\: . 
\label{scs5}
\eneq
\noindent

As one sees from Eqs.(\ref{scs1},\ref{scs2}), 
introducing the contacts does not affect at all the boundary condition
of $\xi ( x , t )$ at the inner boundary ($x = 0$). The boundary conditions
for $X ( x , t )$, instead, explicitly depend on the applied voltage bias, 
whose effect may be accounted for by including in the total Hamiltonian a 
pertinent "source" term.

At the weakly coupled fixed point, biasing the left-hand lead at a dc-voltage 
$V/2$ and the right-hand lead at a voltage $-V/2$, corresponds to adding to 
the SQUID's Hamiltonian a voltage dependent
term given by

\beq
H_V = - \frac{ g \sqrt{2} e^* V }{ 4 \pi}  \int_{-L}^L \: d x \: 
\left[ \frac{ \partial \Phi_R ( x ) }{ \partial t} - 
\frac{ \partial \Phi_L ( x ) }{ \partial t} \right] 
= - \frac{ g  e^* V }{ 2 \pi}  \int_{0}^L \: d x \: 
 \frac{ \partial X ( x ) }{ \partial t}
\:\:\:\: , 
\label{condad3}
\eneq
\noindent
which is equivalent to a shift of $ \Phi (x , t )$, which is linear in time,
 namely, the field $\Phi ( x , t )$ should be replaced by 
$ \Phi ( x , t ) -   e^* V t $.
From  Eqs.(\ref{scs1},\ref{scs2},\ref{condad3}) one gets that also 
$\Phi_{u/d} ( x , t )$ must be shifted as

\beq
\Phi_{u/d} ( x , t ) \to \Phi_{u / d }  ( x , t ) -  
\frac{  e^* V  }{2  } t
\:\:\:\: . 
\label{scs8.0}
\eneq
\noindent

As a consequence, one sees that $X ( x , t )$ takes the additional
contribution, given by 

\beq
X ( x , t ) \to X ( x , t ) -  
 e^* V  t   
\:\:\:\: ,
\label{scs9.0}
\eneq

which introduces in  the boundary interaction Hamiltonian an explicit 
dependence on $\tau$, given by

\[
H_{dc , J} \to H_{dc , J} ( i \tau ) = -  \bar{E}_{J,1}
 \cos \left[ X ( i \tau ) +
 \xi ( i \tau ) + \frac{ \varphi}{2} + i  e^* V  \tau  \right]
\]

\beq
-  \bar{E}_{J,2} \cos \left[ X ( i \tau ) - \xi (i \tau ) - 
\frac{ \varphi}{2} +  i e^* V \tau  \right]
\:\:\:\: ; 
\label{condad5bis}
\eneq
\noindent

at finite $V$, the current 
operators  across the two junctions are then given by

\beq
 j_1 ( i \tau )  = 
e^*  \bar{E}_{J,1} \sin \left[ X ( i \tau ) + \xi ( i \tau ) + \frac{ \varphi}{2}  + i 
 e^* V  \tau   \right]
\:\:\:\: , 
\label{condad4}
\eneq
\noindent

and 

\beq
 j_2 ( i \tau ) = e^*\bar{E}_{J,2}  \sin \left[ X ( i \tau ) - \xi ( i \tau ) +
\frac{ \varphi}{2} - i   e^* V \tau   \right]
\:\:\:\: . 
\label{condad5}
\eneq
\noindent

As in the previous section, the current across the two junctions is 
computed from a perturbative expansion in 
$\bar{E}_{J , 1} , \bar{E}_{J,2}$. One obtains the following results

\begin{itemize}

\item Current across the junction 1: $I_1$

\[
I_1 = I_{1,1} + I_{1 , 2 } 
\:\:\:\: . 
\]
\noindent

$I_{1,1}$ is given by 

\begin{enumerate}

\item

\[
I_{1,1} =  e^* (\bar{E}_{J,1} )^2 \int_0^\beta \: d \tau' \: \langle 
{\bf T}_\tau \{   : \sin [ X ( i \tau ) + \xi ( i \tau ) +
 \frac{ \varphi}{2}  + i
e^* V \tau ]: \times
\]

\[
 : \cos [ X ( i \tau' ) + \xi ( i \tau' ) + \frac{ \varphi}{2}  + i 
e^* V \tau' ]: \} \rangle 
= \frac{e^*}{2} ( \bar{E}_{J,1} )^2  \: \int_0^\infty \: d \tau \:
\biggl\{ \frac{ \sinh \left[ e^* V \tau \right] 
\ }{  \left[ 2 
\sinh \left( \frac{\vpl \pi}{2 L } \tau \right) \right]^\frac{2}{g} } \biggr\}
\:\:\:\: ,
\]

which, for $4 L e^* V / ( \vpl \pi ) \gg 1$ yields

\beq
I_{1,1} \approx  \frac{ 2 e^* (\bar{E}_{J,1} )^2 L}{\pi \vpl } \sin 
\left[ \frac{  \pi}{
g} \right] \Gamma [ 1 - \frac{ 2 }{g} ]  
\left[ \frac{ 4 L e^* V}{ \pi \vpl  } \right]^{  \frac{2}{g} - 1 }
\:\:\:\: .
\label{lia1}
\eneq
\noindent

\item

\[
I_{1,2} =  e^* \bar{E}_{J,1} \bar{E}_{J,2}  \int_0^\beta \: d \tau' \: \langle 
{\bf T}_\tau \{   : \sin [ X ( i \tau ) + \xi ( i \tau ) + \frac{ \varphi}{2} +
e^* V \tau ]:  
\times
\]

\[
: \cos [ X (i \tau' ) - \xi ( i \tau' ) - \frac{ \varphi}{2} +
e^* V \tau' ]: \} \rangle 
\]

\[
\approx  e^* \bar{E}_{J,1} \bar{E}_{J,2} \int_0^\infty \: d \tau \:  \biggl\{ 
\frac{ e^{ e^* V \tau + i \varphi}}{  \left[ \cosh \left( \frac{\pi\vpl}{4 L } 
\tau \right)  \right]^\frac{2}{g} } - \frac{
 e^{ - e^* V \tau -  i \varphi} }{ \left[ \cosh \left( \frac{\pi\vpl}{4 L } 
\tau \right)  \right]^\frac{2}{g} } \biggr\}
\]

\beq
= \frac{ 4 e^* \bar{E}_1\bar{E}_{J,2} L }{ \pi \vpl} \biggl\{ e^{ i \varphi } {\bf J} [ a ] 
- e^{ - i \varphi} {\bf J} [ - a ] \biggr\} 
\:\:\:\: ,
\label{jo1}
\eneq
\noindent
with

\beq
{\bf J} [ a ] = \int_0^\infty \: d x \: 
\frac{ e^{ a x } }{ \cosh^\frac{2}{g} ( x )}
\:\:\:\: ,
\label{jo2}
\eneq
\noindent

where $a = 4 e^* V L / ( \vpl \pi )$ and 

\beq
{\bf J} [ a ] = 2^{ \frac{2}{g} - 1} 
e^{ - i \pi \left( \frac{a}{2} + \frac{1}{g}  \right)  } 
B \left[ - 1 ; - \left( 
\frac{a}{2}  -  \frac{1}{g} \right) ; 1 - \frac{2}{g} \right]
\:\:\:\: , 
\label{jo3}
\eneq
\noindent

where the incomplete - Beta function is defined as
$B [ z , \alpha , \beta ] = \int_0^z \: d t \: 
t^{ \alpha - 1 } \: ( 1 - t )^{\beta - 1 }$

For large $a$ and $1 - 2/g > 0$, one may consider the asymptotic 
expansion of the incomplete- Beta function  \cite{abram},

\beq
 B \left[ - 1 ; - \left( 
\frac{a}{2} + \frac{1}{g} \right) ; 1 - \frac{2}{g} \right] 
\approx  i \exp \left[ - i \frac{\pi}{2} a - i \frac{\pi}{2} 
\left( \frac{2}{g} - 1 \right)  \right] a^{ \frac{2}{g} - 1 }
\Gamma [ \frac{2}{g} - 1 ] 
\:\:\:\: ,
\label{jo7}
\eneq
\noindent

from which 

\[
I_{ 1 , 2 } \approx \frac{ 2 e^* \bar{E}_{J,1} \bar{E}_{J,2} L}{ \pi \vpl } 
\Gamma [1- \frac{2}{g} ] 
\: \biggl\{ \cos ( \varphi ) \cos \left[ \frac{  2 e^* V L}{   \vpl  }
+ \frac{ \pi }{g} \right]
\]

\beq 
 + \sin ( \varphi ) \sin \left[ 
\frac{  2 e^* V L}{   \vpl  } + \frac{ \pi }{g} \right]
\biggr\} \left[ \frac{ 4 L e^* V}{ \pi \vpl } \right]^{ \frac{2}{g} - 1 } 
\:\:\:\: . 
\label{jo8}
\eneq
\noindent

\end{enumerate}

\item Current across the junction 2: $I_2$

\[
I_2 = I_{2,1} + I_{2 , 2 } 
\:\:\:\: , 
\]
\noindent

The two contributions are computed below:

\begin{enumerate}

\item

\[
I_{2,2} =  e^* (\bar{E}_{J,2} )^2 \int_0^\beta \: d \tau' \: \langle 
{\bf T}_\tau \{   : \sin [ X (i \tau ) + \xi ( i \tau ) + 
\frac{ \varphi}{2}  - i
e^* V \tau ]:  
\times
\]

\beq
: \cos [ X ( i \tau' ) + \xi ( i \tau' ) + \frac{ \varphi}{2}  - i 
e^* V \tau' ]: \} \rangle 
\approx  \frac{ 2 e^* (\bar{E}_{J,2} )^2 L}{\pi \vpl } \sin 
\left[ \frac{  \pi}{
g} \right] \Gamma [ 1 - \frac{ 2 }{g} ]  
\left[ \frac{ 4 L e^* V}{ \pi \vpl  } \right]^{  \frac{2}{g} - 1 }
\:\:\:\: .
\label{lia1.1}
\eneq
\noindent

\item

\[
I_{2,1} =  e^* \bar{E}_{J,1} \bar{E}_{J,2}  \int_0^\beta \: d \tau' \: \langle 
{\bf T}_\tau \{   : \sin [ X ( i \tau ) + \xi ( i \tau ) - \frac{ \varphi}{2} +
e^* V \tau ]: 
\times
\]

\[
 : \cos [ X ( i \tau' ) - \xi (i  \tau' ) + \frac{ \varphi}{2} +
e^* V \tau' ]: \} \rangle 
\]

\[
 \approx \frac{ 4 e^* \bar{E}_1\bar{E}_{J,2} L}{ \pi \vpl } 
\Gamma [1- \frac{2}{g} ] 
\: \biggl\{ \cos ( \varphi ) \cos \left[ \frac{  2 e^* V L}{   \vpl  }
+ \frac{ \pi }{g} \right]
\]

\beq 
 + \sin ( \varphi ) \sin \left[ 
\frac{  2 e^* V L}{   \vpl  } + \frac{ \pi }{g} \right]
\biggr\} \left[ \frac{ 4 L e^* V}{ \pi \vpl } \right]^{ \frac{2}{g} - 1 } 
\:\:\:\: . 
\label{jo8.1}
\eneq
\noindent

\end{enumerate}
 
\end{itemize}

The net transport current across the dc-SQUID 
is obtained as the sum of the currents across the
two junctions, that is

\[
I = I_1 + I_2 = \frac{ 2 e^* L}{\pi u } 
 \Gamma [ 1 - \frac{ 2}{g} ]  \biggl\{  [ (\bar{E}_{J,1})^2 
+(\bar{E}_{J,2} )^2 ]
 \sin \left[ \frac{  \pi}{g} \right]
\]

\beq
+  \bar{E}_{J,1} \bar{E}_{J,2} 
 \cos ( \varphi ) \cos \left[ \frac{  2 e^* V L}{   \vpl  }
+ \frac{ \pi }{g} \right] \biggr\}
\left[ \frac{ 4 L e^* V}{ \pi u  } \right]^{  \frac{2}{g} - 1 }
\:\:\:\: . 
\label{trdcsq1}
\eneq
\noindent

Eq.(\ref{trdcsq1}) contains an "interference" contribution 
(proportional to $\cos ( \varphi )$), representing the cumulative effect
of a bias voltage applied to junction 1 on the current flowing across 
junction 2, and of a bias voltage  applied to junction 2 on the current 
flowing across  junction 1. This is a remarkable result of the application 
of the boundary sine Gordon techniques to SQUID devices. Treating the
plasmon modes as modes of the dynamical fields, allows for an explicit
computation of the interference contributions to the dc-current.

The circulating current across the dc-SQUID, $J$, may be computed as the 
difference between the currents flowing across the two junctions, $I_J = 
I_1 - I_2$. The result is

\beq 
J =  \frac{ 4 e^* \bar{E}_{J,1} \bar{E}_{J,2}  L}{ \pi \vpl } 
\Gamma [1- \frac{2}{g} ]  \sin ( \varphi ) \sin \left[ 
\frac{  2 e^* V L}{   \vpl  } + \frac{ \pi }{g} \right]
\biggr\} \left[ \frac{ 4 L e^* V}{ \pi \vpl } \right]^{ \frac{2}{g} - 1 } 
\:\:\:\: ,
\label{jo8.10}
\eneq
\noindent
which, again, comes from pertinently taking into account correlation
functions of the plasmon modes computed using the boundary sine Gordon 
theory.

\subsection{The dc-SQUID at the strongly coupled fixed point}

At the strongly coupled fixed point,  as long as one neglects
phase slips at the junctions, one should not  expect any 
 resistance across the 
junctions.  As a consequence, the external voltage $V$ is added by biasing 
the left chiral mode of the plasmon field at each lead with respect to the 
right chiral mode, with opposite biases in the two contacts.
As discussed in section (7.2) for the chain, this is accounted for by 
shifting the field $\Phi ( x , t )$ according to

\beq
\Phi ( x , t ) \to \Phi ( x , t ) -  \frac{  e^* V  }{\vpl} x
\:\:\:\: .
\label{scs7}
\eneq
\noindent

Due to  Eqs.(\ref{scs1},\ref{scs2}, \ref{scs7}), one has that, at finite 
$V$, also $\Phi_{u/d} ( x , t )$ must be shifted as

\beq
\Phi_{u/d} ( x , t ) \to \Phi_{u / d }  ( x , t ) -  
\frac{  e^* V  }{2 \vpl } x
\:\:\:\: . 
\label{scs8}
\eneq
\noindent

Eqs.(\ref{scs5}) imply that  

\beq
X ( x , t ) \to X ( x , t ) -  
\frac{  e^* V  }{ \vpl } x \:\:\: ; \:\:
\xi ( x , t ) \to \xi ( x , t )  
\:\:\:\: , 
\label{scs9}
\eneq
\noindent
which is equivalent to adding to the  total Hamiltonian a voltage dependent
contribution given by

\beq
\tilde{H}_V = - \frac{e^* V g }{   \pi} \int_0^L \: d x \: \frac{ \partial 
X ( x )}{ \partial x} = 
 - \frac{e^* V g }{  2 \pi \vpl } \int_0^L \: d x \: \frac{ \partial 
\Theta_X ( x )}{ \partial \tau}
\:\:\:\: .
\label{scs11}
\eneq
\noindent

The current operators across the two junctions are now written as

\beq
j_{1 / 2 }  ( i \tau ) = - i \frac{e^*}{2 \pi} \left[ \frac{ \partial 
\Theta_X ( i \tau )}{ \partial \tau } \pm \frac{ \partial 
\Theta_\xi ( i \tau )}{ \partial \tau } \right] = j_X ( i \tau ) \pm 
j_\xi ( i \tau )
\:\:\:\: . 
\label{scs12}
\eneq
\noindent

An useful representation may be derived by 
 adding to the total Euclidean action $S_E [ \Theta_X , \Theta_\xi , a_X , 
a_\xi ]$ the source term  given by

\beq
\delta S_E^{\rm Source} = \frac{i e^*}{2 \pi L } \int_0^\beta \: d \tau \:
\int_0^L \: d x \: \left[ a_X ( \tau ) \frac{ \partial 
\Theta_X ( i \tau )}{ \partial \tau } +  a_\xi ( \tau ) \frac{ \partial 
\Theta_\xi ( i \tau )}{ \partial \tau } \right]
\:\:\:\: ,
\label{scs14}
\eneq
\noindent
and by taking the functional derivatives of 
$S_E [ \Theta_X , \Theta_\xi , a_X , a_\xi ]$ with respect to
$ a_X , a_\xi $, after shifting the dual fields according to

\beq
\Theta_X ( x , i \tau ) \to \Theta_X ( x , i \tau ) -  
2 i \frac{ e^* g \vpl }{L} 
A_X ( \tau ) - i g e^* V \tau  
\:\:\:\: , 
\eneq
\noindent

and

\beq
\Theta_\xi ( x ,i  \tau ) \to \Theta_\xi ( x ,i  \tau ) - 
 2 i \frac{ e^* g \vpl }{L}  A_\xi ( \tau ) + i  \frac{\varphi}{2 L } \tau
\:\:\:\: ,
\label{scs16} 
\eneq
\noindent
with $A'_X ( \tau ) = a_X ( \tau )$, $A'_\xi ( \tau  ) = a_\xi (\tau )$.

The Euclidean action in the dual representation is then given by

\[
\tilde{S}_E = 
 \sum_{ a = X , \xi } \: \frac{1}{ 4 \pi ( 2g) } \int_0^\beta \: d \tau \:
\int_0^L \: d x \: \left[ \frac{1}{\vpl} \left( \frac{ \partial \Theta_a ( x , 
i \tau )}{ \partial \tau } \right)^2 + 
\vpl \left( \frac{ \partial \Theta_a ( x , 
i \tau )}{ \partial x } \right)^2 \right] 
\]

\[
- 2 Y_1 \int_0^\beta \: d \tau \: \cos \left[ \frac{1}{2} ( \Theta_X ( i \tau )
+ \Theta_\xi ( i \tau ) ) + 
i \left[ g e^* V  - \vpl \frac{ \varphi}{4 L } \right] \tau  + 
i \frac{ 2 e^* g \vpl }{L} 
[ A_\xi ( \tau ) + A_X ( \tau ) ] \right]
\]

\[
- 2 Y_2 \int_0^\beta \: d \tau \: \cos \left[ \frac{1}{2} (-  \Theta_X 
( i \tau ) + \Theta_\xi (i  \tau ) ) -  
i \left[ g e^* V  + \vpl \frac{ \varphi}{4 L } \right] \tau + 
i \frac{ 2 e^* g \vpl }{L} [  A_\xi ( \tau ) -  A_X ( \tau ) ] \right]
\]

\beq
+ \frac{1}{ 4 \pi (2g) } \int_0^\beta \: d \tau \: \int_0^L \: d x \: 
\biggl\{ \left[ 2 g e^* V + 2 e^* g \vpl \frac{a_X ( \tau )}{L} \right]^2 
+  \left[2 e^* g \vpl \frac{a_\xi ( \tau )}{L}  + \vpl \frac{ \varphi}{ 4 L } 
\right]^2 \biggr\}
\:\:\:\: . 
\label{scs17}
\eneq
\noindent

As a result of Eq.(\ref{scs17}), one  finds that the current operators 
$j_X$, $j_\xi$ may be written as 

\[
j_X ( i \tau ) = \frac{g ( e^* )^2 }{ \pi  } V +
 4 i \frac{ e^* g \vpl Y_1}{L} \:
 \sin \left[ \frac{1}{2} ( \Theta_X ( i \tau )
+  \Theta_\xi ( i \tau ) ) + 
i \left[ g e^* V  + \vpl \frac{ \varphi}{4 L } \right]  \tau  \right]
\]

\beq
- 
2 i  \frac{ e^* g \vpl Y_2}{L}  
 \sin \left[ \frac{1}{2} (-  \Theta_X ( i \tau )
+ \Theta_\xi ( i \tau ) ) - 
i \left[ g e^* V  + \vpl \frac{ \varphi}{4 L } \right]  \tau  \right]
\:\:\:\: ,
\label{scs18}
\eneq
\noindent

and by

\[
j_\xi ( i \tau ) =   \frac{e^*  g \vpl }{ 4 \pi L }  \varphi
 + 4 i \frac{ e^* g \vpl Y_1}{L} \:
 \sin \left[ \frac{1}{2} ( \Theta_X ( i \tau )
+  \Theta_\xi ( i \tau ) ) + 
i \left[ g e^* V  - \vpl \frac{ \varphi}{4 L } \right]  \tau  \right]
\]

\beq
+
4 i  \frac{ e^* g u Y_2}{L}  
 \sin \left[ \frac{1}{2} (-  \Theta_X ( i \tau )
+ \Theta_\xi ( i \tau ) ) - 
i \left[ g e^* V  + \vpl \frac{ \varphi}{4 L } \right]  \tau  \right]
\:\:\:\: ,
\label{scs19}
\eneq
\noindent

From Eqs.(\ref{scs18},\ref{scs19}), one sees that,  to the lowest order in the
phase slip contribution ($Y_1 = Y_2 = 0 $), the stationary component of the
circulating current reduces to 
the value of the Josephson current at zero 
voltage, $ \frac{e^*  g u }{ 4 \pi L }  \varphi$,
while the  conduction current is given by  $\frac{g ( e^* )^2 }{ \pi  } V$,
just as for a single chain at the strongly coupled fixed point. 

Higher-order corrections, including the effects of phase slips, may be
computed following the same steps used in the previous section. 
The result is

\[
\delta j_X  = - \frac{4 e^*  g (Y_1)^2 }{\pi } 
\biggl\{ \Gamma [ 1 - 2 g ] \biggl[ \frac{ \Gamma [ - \frac{a_-}{2} + 
g ] }{ \Gamma [ 1   - \frac{a_-}{2} - g ] }
-  \frac{ \Gamma [  \frac{a_-}{2} + 
g ] }{ \Gamma [ 1 + \frac{a_-}{2} - g ] } \biggr] 
\biggr\}
\]

\beq
+   \frac{4 e^*  g (Y_2)^2 }{\pi } 
\biggl\{ \Gamma [ 1 - 2 g ] \biggl[ \frac{ \Gamma [  \frac{a_+}{2} + 
g ] }{ \Gamma [ 1   + \frac{a_+}{2} - g ] }
-  \frac{ \Gamma [ - \frac{a_+}{2} + 
g ] }{ \Gamma [ 1 - \frac{a_+}{2} - g ] } \biggr] 
\biggr\}
\:\:\:\: , 
\label{scs22}
\eneq
\noindent

with

\beq
a_\mp = \frac{ 2 g e^* V L }{\pi u } \mp \frac{ u \varphi}{\pi} 
\:\:\:\: , 
\label{scs23}
\eneq
\noindent

and by

\[
\delta j_\xi  = - \frac{4 e^*  g (Y_1)^2 }{\pi } 
\biggl\{ \Gamma [ 1 - 2 g ] \biggl[ \frac{ \Gamma [ - \frac{a_-}{2} + 
g ] }{ \Gamma [ 1   - \frac{a_-}{2} - g ] }
-  \frac{ \Gamma [  \frac{a_-}{2} + 
g ] }{ \Gamma [ 1 + \frac{a_-}{2} - g ] } \biggr] 
\biggr\}
\]

\beq
-   \frac{4 e^*  g (Y_2)^2 }{\pi } 
\biggl\{ \Gamma [ 1 - 2 g ] \biggl[ \frac{ \Gamma [  \frac{a_+}{2} + 
g ] }{ \Gamma [ 1   + \frac{a_+}{2} - g ] }
-  \frac{ \Gamma [  - \frac{a_+}{2} + 
g ] }{ \Gamma [ 1 - \frac{a_+}{2} - g ] } \biggr] 
\biggr\}
\:\:\:\: , 
\label{scs22bis}
\eneq
\noindent

For $\varphi \sim 0$ and for $| a_\pm | \ll 1$, one may expand 
 Eqs.(\ref{scs22},\ref{scs22bis}) to the first order in $a_\pm$.
In the symmetric case ($Y_1 = Y_2$), this provides a correction to the slope
of both the dc-current and of the stationary component of the 
Josephson current, so that one gets

\beq
\delta 
j_X \approx \frac{(e^* )^2 g}{2 \pi} \biggl\{ 1 + \frac{16 Y^2  g L}{\pi u }
C \Gamma [ 1 - 2 g ] \biggr\} V
\;\;\;\; , 
\label{scs221}
\eneq
\noindent

and

\beq
\delta 
j_\xi \approx \frac{e^* g \vpl}{4 \pi L} 
\biggl\{ 1 + \frac{16 Y^2  g L}{\pi u }
C \Gamma [ 1 - 2 g ] \biggr\} \varphi
\;\;\;\; , 
\label{scs221.b}
\eneq
\noindent

with

\beq
C = \frac{ \partial }{ \partial x} \biggl\{ \frac{ \Gamma [ z + g ]}{ 
\Gamma [ 1 - z + g ]} - \frac{ \Gamma [ z + g ]}{ \Gamma [ 1 + z - g ]} 
\biggr\}_{ z = 0 }
\:\:\:\: . 
\eneq
\noindent

Eqs.(\ref{scs221},\ref{scs221.b}) show in a rather simple context the
way in which relevant perturbation may affect both the Josephson and the
normal currents across the device.

\section{Comments and conclusions}

We provided a framework where two-boundary sine Gordon models may be
used  to investigate the relevant  nonperturbative features,
occurring in the renormalization of the Josephson couplings
$E_{J,i}$ of a multi-junction SQUID. As pointed out in
Ref.\cite{glazhek}, this is needed since -for large fluctuations of
the phase of the order parameter- the effective boundary conditions
on the plasmon fields depend on the Josephson energy of the
junction. Our analysis shows  that, when $g>1$ and $L>L_*$,
the Josephson couplings affect the dynamics of the plasmon field
only as a boundary term, while, for $L<L_*$, they can be regarded as 
quantum impurities in a superconducting loop. For $g<1$, instead, 
the Josephson couplings have to be always regarded as quantum impurities.

The boundary field theory approach to SQUID devices proposed in this
paper turns out to be very powerful in providing a systematic procedure 
for computing not only the Josephson currents (and the leading corrections 
induced by the relevant perturbations), but also the dc-currents flowing in
a SQUID device, due to externally applied voltages. For pertinent choices
of the infrared cutoff, the results of the boundary field theory
well reproduce what has been obtained in Ref.\cite{zwergerkane,glazhek},
allowing, in addition, to appreciate new and closer
connections between the theories of Josephson superconducting devices, quantum
brownian motion, fermionic quantum wires, and quantum impurity
problems. Furthermore, for dc-SQUID devices, our analysis points out
the existence of new remarkable 
interference effects between the currents flowing through each junction.

We investigated here SQUID-devices with one or two junctions. For these
systems, we have found two stable regimes characterized by stable
fixed points of the pertinent RG equations, driven by the strength
of the bare Josephson energy. It would be interesting to ascertain if 
superconducting loops with more than two junctions may exhibit 
 new (in addition to the by now well known weakly and strongly coupled
fixed points) renormalization group fixed points at a finite Josephson
coupling. Furthermore, it could be instructive to investigate the
properties of a multi-junction SQUID device, for which 
$g=1$, since the boundary sine-Gordon model provides an exact solution for
this value of the Luttinger parameter. 
We feel that multi-junction superconducting loops may become
an interesting laboratory for testing the physical properties of
field theories describing interacting Luttinger liquids with
pertinent boundary interactions.

\label{}

\vspace{1cm}

{\bf Acknowledgments}

\vspace{0.5cm}

We thank Ian Affleck, Leonid Glazman,  Hubert Saleur and Gordon W. Semenoff 
for inspiring discussions and useful correspondence at various stages of this
project. We benefited  from conversations with Maria Cristina
Diamantini, Gianluca Grignani, Francesco P. Mancini, Arianna Montorsi, 
Mario Rasetti, Arturo Tagliacozzo and Andrea Trombettoni.
This work has been partly supported by the MIUR National Project
{\it Josephson Networks for Quantum Coherence and Information}
(grant No.2004027555).

 \appendix

\section{Scaling properties of boundary interaction strengths}

In this appendix we shall derive the scaling laws for boundary
interaction operators.

Let us consider a one-dimensional system described by the Euclidean action

\beq
S = S_0 - \sum_{ \ell = 1}^M \lambda_\ell \int_0^\beta \: d \tau \:
\Phi_\ell ( \tau ) \:\:\: \: ,
\label{ap1.1}
\eneq
\noindent

where $S_0$, the action for $K$ independent massless Klein-Gordon
fields, $\{ \phi_j \}$, with pertinent boundary conditions, is given
by:

\beq
S_0 = \frac{g}{ 4 \pi} \sum_{ j = 1 }^K \int_0^\beta \: d \tau \:
\int_0^L \: d x \: \left[ \frac{1}{\vpl} \left( \frac{ \partial \phi_j}{
\partial \tau } \right)^2 +  \vpl \left( \frac{ \partial \phi_j}{
\partial \tau } \right)^2 \right]
\:\:\: .
\label{ap1.2}
\eneq
\noindent

The boundary operators $\{ \Phi_\ell ( \tau ) \}$ are functionals of
the fields $\{ \phi_j \}$ with scaling dimension $h_\ell$.  As a
consequence, for $\tau \to \tau'$, one gets

\beq G_\ell ( \tau - \tau' ) = \langle \Phi_\ell ( \tau )
\Phi_{\ell'} ( \tau' ) \rangle_0 \propto \delta_{ \ell , \ell^{'}}
\frac{1}{ [2 \pi  \vpl ( \tau - \tau' ) / L ]^{2 h_\ell } } \:\:\:\:
, \label{ap1.3} \eneq \noindent where $ \langle \ldots \rangle_0$
denotes averaging with respect to the "free" action $S_0$, while $L$
is the "size" of the system.

It is well known that, in order to introduce scale invariant
interaction terms, one needs to define dimensionless coupling
constants, given by
 $g_\ell ( a ) = \lambda_\ell a^{ 1 - h_\ell }$, where
 $\lambda_\ell$
 is the coupling constant appearing in Eq.(\ref{ap1.1}), and $a$ is
 a short distance cutoff. To first order in the
coupling strengths, the renormalization group equations are given by

\beq
\frac{ d g_\ell ( a ) }{ d \ln ( a / a_0 ) } = [ 1 - h_\ell ] g_\ell ( a )
\:\:\:\: .
\label{ap1.7}
\eneq
\noindent

Higher order corrections to the renormalization group equations come
from nontrivial short-distance fusion rules of the  $\Phi_\ell$'s,
which are represented by the O.P.E.'s

\beq
 \Phi_\ell ( \tau )  \Phi_{\ell'} ( \tau' )  \approx_{ \tau' \to \tau }
\sum_k \frac{ C_{\ell \: \ell' \: \ell"} }{ | \tau - \tau' |^{h_\ell
+ h_\ell' - h_\ell" } } \Phi_{\ell"} ( \tau ) \:\:\:\: .
\label{ap1.11} \eneq \noindent

If the O.P.E.'s coefficients $C_{\ell \: \ell' \: \ell"} $ are
different than zero, the scaling equations get contributions which
are of second order in the couplings.  To derive them, one should
start from the partition function

\beq  Z =  Z_* \left \langle \exp \left[ \sum_{\ell = 1}^M g_\ell (
a ) \int_0^\beta \: d \tau \: a^{ - 1 + h_\ell } \Phi_\ell ( \tau )
\right] \right \rangle_0 \:\:\:\: . \label{ap1.12} \eneq \noindent

Upon introducing the cutoffs $a$ and $L$ and by expanding
Eq.(\ref{ap1.12}) up to the third-order in the running coupling
strengths, one gets

\[
\frac{  Z }{ Z_* } \approx 1 + \sum_{ \ell , \ell' = 1}^M g_\ell
g_{\ell'} \int_0^{ L / \vpl } d \: \tau \: \int_0^{ L / \vpl } \: d
\tau' \: a^{ - 1 + h_\ell} a^{ - 1 + h_{\ell'} } \theta ( \tau  -
\tau' - \frac{a}{\vpl} ) \langle  \Phi_\ell ( \tau )
 \Phi_{\ell'} ( \tau' )  \rangle_0
\]

\[
+ \sum_{\ell,\ell',\ell'' = 1}^M g_\ell g_{\ell'} g_{\ell''}
\int_0^\frac{L}{u}  \: d \tau \: \int_0^{ L / \vpl }
\: d \tau' \: \int_0^{ L / \vpl } \: d \tau'' \: a^{ - 1 + h_\ell }
a^{ - 1 + h_{\ell'}} a^{ - 1 + h_{\ell''}} \times
\]

\beq
\theta ( \tau - \tau' -
\frac{a}{\vpl } ) \theta ( \tau' - \tau'' - \frac{a}{\vpl} ) \langle
 \Phi_\ell ( \tau )  \Phi_{\ell'} ( \tau' )  \Phi_{\ell''} ( \tau'' )
\rangle_0  \:\:\:\: .
\label{ap1.13}
\eneq
\noindent

If one rescales the cutoff, $a \to (1+\epsilon ) a $, with $\epsilon
\ll 1$, additional contributions to the coupling constant
renormalization arise, as a result of  configurations where the
arguments of two fields lie between $a $ and $ (1 + \epsilon ) a $.
The ensuing  renormalization is derived by means of the identity

\[
\theta ( \tau - \tau' - \frac{a}{\vpl} ( 1 + \epsilon ) ) \;
\theta ( \tau' - \tau'' -  \frac{a}{\vpl} ( 1 + \epsilon )  )
\approx \theta ( \tau - \tau' - \frac{a}{\vpl} ) \; \theta ( \tau' - \tau'' -
\frac{a}{\vpl} )
\]

\beq
 - \epsilon \frac{a}{\vpl} [ \delta ( \tau - \tau' - \frac{a}{\vpl} )
\theta ( \tau' - \tau'' - \frac{a}{\vpl} ) +
\theta ( \tau - \tau' - \frac{a}{\vpl}
) \delta ( \tau' - \tau'' - \frac{a}{\vpl} ) ] \:\:\:\: .
\label{ap1.14}
\eneq
\noindent

Thus, the third-order contribution to Eq.(\ref{ap1.13}) renomalizes
the second-order term with

\[
 2 \epsilon \sum_{\ell' , \ell''=1}^M   g_{\ell'} g_{\ell''}
\sum_{ \ell }^M [ g_\ell  C_{\ell , \ell' , \ell'' }  ]
\int_0^{ L / \vpl } \: d \tau' \: \int_0^{ L / \vpl } \: d \tau" \:
\theta ( \tau' - \tau" - \frac{a}{\vpl} ) \times
\]

\beq
 a^{ - 1 + h_{\ell'} } a^{ - 1 + h_{\ell"} }
\Phi_{\ell'} ( \tau' )  \Phi_{\ell"} ( \tau" )  \:\:\:\: .
\label{ap1.15} \eneq \noindent By setting $\epsilon = \ln ( a / a_0
)$ in Eq.(\ref{ap1.12}), it is straightforward, but tedious, to
derive explicit expression for the nonlinear terms appearing in the
RG equations for the running coupling strengths; the final result is

\beq
\frac{ d g_\ell ( a ) }{ d \ln ( a / a_0 )  } = [ 1 - h_\ell ] g_\ell ( a )
+  \sum_{ \ell' , \ell'' = 1 }^M C_{\ell , \ell' , \ell''} g_{\ell'} ( a )
g_{\ell''} ( a ) \:\:\:\: .
\label{ap1.16}
\eneq
\noindent

The RG equations in Eq.(\ref{ap1.16}) may be  rewritten as

\beq \frac{d g_\ell}{ d \ln ( a / a_0 ) } - \frac{ \partial C [ \{ g
\} ] }{ \partial g_\ell} = 0
\:\:\:\:,
 \label{ap1.17} \eneq
\noindent where $C  [ \{ g \} ]$ is  given by

\beq
C [ \{ g \} ] = \frac{ 1}{2} \sum_{ \ell = 1}^N [ 1 - h_\ell] g_\ell^2
+ \frac{1}{3} \sum_{ \ell , \ell' , \ell'' = 1 }^M
C_{\ell , \ell' , \ell''} g_\ell g_{\ell'} g_{\ell''} \:\:\:\: .
\label{ap1.18}
\eneq
\noindent

The RG fixed points coincide with the extrema of the function $C [
\{ g \} ]$, that is, with the set of values of $ \{ g \}$, $\{ g^*
\}$, such that

\beq \frac{ \partial C  [ \{ g^* \} ] }{ \partial g_\ell } = 0
\:\:\:\: , \:\: \forall \ell = 1 , \ldots , M \:\:\:\: .
\label{ap1.19} \eneq \noindent Due to its properties, $C$ may also
be identified with the "boundary entropy" of the system
\cite{Afflud}.

We would like to point out that the thermodynamics limit may be
achieved either by sending either $a$ or $L$ to $\infty$. In fact,
sending $a$ to $\infty$ amounts to cut off high momenta
contributions, which amounts to send the size of the system to
$\infty$. This implies that scaling may be realized either using $a
/ a_0$, or $L / L_*$ (as it has been done in this paper), as scaling
parameters.

\section{Boundary vertex operators}

As an example of correlation functions computed with both $\beta$ and $L$
finite, in this appendix we derive  the correlators of the normal-ordered
vertices  $V_{X , \pm 1} ( \tau ) = : \exp \left[ \pm i   X ( i \tau ) 
\right]:$, and $V_{\xi , \pm 1} ( i \tau ) = : \exp \left[ \pm i   
\xi ( i \tau ) \right]:$ introduced in section 3, 
where $X ( x , t )$ obeys
Neumann boundary condions at both boundary $x=0$ and $x=L$, while 
$\xi ( x , t )$ obeys Dirichlet boundary conditions at the inner boundary,
Neumann at the outer boundary. 

The partition function  $ Z_X= {\rm Tr} [ - \beta H_X ]$, where $H_X$ is 
the Hamiltonian for the field $X$, has been derived in Eq.(\ref{bc12}).
The result is 
\beq Z_X = \frac{1}{ \eta ( q ) }\sum_{ m \in Z } \exp \left[ -
\beta \frac{  2 g  \pi \vpl}{L} m^2  \right] \:\:\:\: . \label{apb6b}
\eneq \noindent

The two-vertex correlation function, $G_X ( \tau_1 , \tau_2 )$, is
then given by

\beq G_X ( \tau_1 , \tau_2 ) = \frac{1}{  Z_X} {\rm Tr} [ T_\tau \{
: e^{ -  i  X  ( i \tau_1 )} :
 : e^{    i  X  ( i \tau_2 )} : \} e^{ - \beta H_X} ]
\:\:\:\: .
\label{apb11}
\eneq
\noindent

As it happens for the partition function, also $G_X$ is factorized
into a contribution from the oscillatory modes, times a contribution
from the zero modes. The trace over the oscillatory modes yields

\beq
{\bf T}_{\rm osc} = \prod_{ n = 1}^\infty {\bf T}_n \:\:\:\: ,
\label{apb12}
\eneq
\noindent

where

\[
{\bf T}_n = {\rm Tr} {\bf T}_\tau \biggl\{ \exp \left[ i \sqrt{\frac{1}{g}} 
\frac{ 
a ( - k_n ) }{ n} e^{ \vpl k_n \tau_1 } \right]
\exp \left[ - i \sqrt{\frac{1}{g}}    \frac{ a  (  k_n ) }{ n} e^{-
\vpl k_n \tau_1 } \right] \times
\]

\beq \exp \left[ i \sqrt{ \frac{2}{g} } \frac{ a_X ( - k_n ) }{ n} e^{
\vpl k_n \tau_2 } \right] \exp \left[  i \sqrt{ \frac{2}{g} } \frac{
a_X  (  k_n ) }{ n} e^{-  \vpl k_n \tau_2 } \right] q^{ a_X  ( - k_n )
a_X ( k_n ) } \:\:\:\: .\label{apb13} \eneq \noindent The trace in
Eq.(\ref{apb13})  may be computed by resorting to a coherent state
decomposition, which uses the basis of coherent states $ | x_n
\rangle$, defined by

\beq a_X ( k_n ) | x_n \rangle = \sqrt{n} x_n | x_n \rangle
\Rightarrow | x_n \rangle = e^{ - \frac{ | \vec{\xi}_n |^2}{2} } e^{
\frac{x_n  a_X ( - k_n ) }{ n} } | 0 \rangle \:\:\:\: ,
\label{apb14} \eneq \noindent and the decomposition of the identity
given by

\beq
{\bf I} = \left( \frac{1}{ \pi} \right)
\int \: d x_n \: e^{ - \frac{ ( x_n )^2}{2 }}
| x_n \rangle \langle x_n | \:\:\:\: .
\label{apb15}
\eneq
\noindent

The derivation is tedious, but straightforward \cite{gsw}. The final
result is

\[
{\bf T}_n = \left[ \frac{1}{ 1 - q^n} \right]^2
\exp \left[ - \frac{ 4}{g n} \left(
\frac{q^n}{ 1 - q^n} \right) \right] \times
\]

\beq \exp \left\{ \left( \frac{ 2}{ g ( 1 - q^n) } \right) \frac{
e^{ - \vpl k_n | \tau_1 - \tau_2 | } }{n} + \left( \frac{ 2 q^n}{ g
( 1 - q^n ) } \right) \frac{ e^{  \vpl k_n | \tau_1 - \tau_2 | }
}{n} \right\} \:\:\:\: . \label{apb16} \eneq \noindent

As a consequence, one gets

\[
{\bf T}_{\rm osc} = \prod_{n = 1}^\infty {\bf T}_n =
\frac{1}{ \eta ( q ) }  \exp \left[ - \frac{4}{g}
\sum_{ n = 1}^\infty \frac{ q^n}{
n ( 1 - q^n ) } \right] \times
\]

\beq
\exp \left[  \frac{ 1}{g} \sum_{ n = 1}^\infty \frac{ ( e^{ - k_n \vpl
 | \tau_1 -\tau_2 | } )^n}{ ( 1 - q^n ) } +
 \frac{ 1}{g} \sum_{ n = 1}^\infty \frac{ q^n ( e^{ - k_n \vpl   | \tau_1 -
\tau_2 | } )^n}{ ( 1 - q^n ) } \right] \:\:\:\: . \label{apb17}
\eneq

Using the Jacobi's triple product identity \cite{gsw}, one finally
obtains

\beq
\prod_{n = 1}^\infty {\bf T}_n = \frac{1}{ \eta ( q ) }
\left\{  
\frac{ \theta_1 \left( \frac{ i \vpl ( \tau_1 - \tau_2 )}{2 L} | i \frac{  \vpl
\beta}{L} \right) }{ \theta^{'}_1 \left( 0 |  i \frac{   \vpl
\beta}{L} \right) } \right\}^{ -  \frac{1}{g} }
\:\:\:\: ,
\label{apb19}
\eneq
\noindent
where $\theta_1 ( \zeta | \tau )$ is Jacobi's $\theta_1$ elliptic function,
defined as

\beq \theta_1 ( \zeta | \tau ) = - i \sum_{ n \in Z} ( - 1 )^n e^{
\pi \tau \left( n + \frac{1}{2} \right)^2 } e^{ 2 i \pi \zeta \left ( n
+ \frac{1}{2} \right)} \:\:\:\: . \label{apb20} \eneq \noindent

By tracing over the zero mode and applying Poisson's resummation
formula, one eventually gets

\beq
G_X ( \tau_1 , \tau_2 )  = 
\theta_3 \left( \frac{ \tau u}{2 g L } \biggr| i \frac{ \beta u}{2 g L } 
\right) 
 \left[   \frac{ \theta_1 \left(
i \vpl \frac{ ( \tau_1 - \tau_2 ) }{2 L} |  
i \frac{   \vpl \beta}{L} \right) }{
\theta^{'}_1 \left( 0 |  i \frac{   \vpl \beta}{2 L} \right) }
\right]^{ -  \frac{ 2}{g} }  \:\:\:\: .
\label{eqb21}
\eneq
\noindent
where the $\theta_3$-function is defined as

\beq
\theta_3 ( \zeta | \tau ) = \sum_{ n \in Z} e^{ i \pi \tau \frac{n^2}{2} }
e^{ 2 \pi n \zeta }
\:\:\:\: . 
\label{thetaf44}
\eneq
\noindent

By performing similar computations, one obtains

\[
G_\xi ( \tau_1 , \tau_2 ) = \frac{1}{  Z_X} {\rm Tr} [ T_\tau \{
: e^{ -  i  \xi  ( i \tau_1 )} :
 : e^{    i  \xi ( i \tau_2 )} : \} e^{ - \beta H_\xi} ] 
=
\]

\beq
\theta_3 \left( \frac{ \tau u}{2 g L } \biggr| i \frac{ \beta u}{2 g L } 
\right)  \left[   \frac{ \theta_4 \left(
i \vpl \frac{ ( \tau_1 - \tau_2 ) }{2 L} |  
i \frac{   \vpl \beta}{L} \right) }{
\theta^{'}_1 \left( 0 |  i \frac{   \vpl \beta}{2 L} \right) }
\right]^{ -  \frac{ 2}{g} }  \:\:\:\: .
\label{eqb21.1}
\eneq
\noindent
where the $\theta_4$-function is defined as

\beq
\theta_4 ( \zeta | \tau ) = \sum_{ n \in Z} (-1)^n e^{i \pi \tau 
\frac{n^2}{2} } e^{ 2 \pi i  n  \zeta}
\:\:\:\: . 
\label{thetaf4}
\eneq

When taking the limit $\beta \to \infty$, the correlators in 
Eqs.(\ref{eqb21},\ref{eqb21.1}) give the ones that we have used in the paper.

\end{document}